\documentclass[fleqn,usenatbib]{mnras}
\usepackage{newtxtext,newtxmath}
\usepackage[T1]{fontenc}
\usepackage{adjustbox}
\usepackage{xcolor}

\DeclareRobustCommand{\VAN}[3]{#2}
\let\VANthebibliography\thebibliography
\def\thebibliography{\DeclareRobustCommand{\VAN}[3]{##3}\VANthebibliography}

\usepackage{graphicx}	
\usepackage{amsmath}	
\usepackage{bm}
\usepackage{xcolor}
\definecolor{texas}{HTML}{BF5700}

\definecolor{navy}{HTML}{0047AB}

\definecolor{ucilblue}{HTML}{6aa2b8}

\newcommand{\msun}{\mathrm{M}_{\odot}}
\newcommand{\kms}{{\rm km\,s}^{-1}}

\title[GCs in dwarf galaxies]{Formation of proto-globular cluster candidates in cosmological simulations of dwarf galaxies at $\bm{z>4}$}

\author[Sameie et al.]{Omid Sameie,$^{1}$\thanks{E-mail: sameie@utexas.edu}
Michael Boylan-Kolchin,$^{1}$\thanks{E-mail: mbk@astro.as.utexas.edu}
Philip F. Hopkins,$^{2}$ 
Andrew Wetzel,$^{3}$  
Xiangcheng Ma,$^{4}$\newauthor
James S. Bullock,$^{5}$
Kareem El-Badry,$^{6,7}$ 
Eliot Quataert,$^8$ 
Jenna Samuel,$^1$
Anna T.~P. Schauer,$^1$ \newauthor
Daniel R. Weisz,$^4$
\\
$^{1}$ Department of Astronomy, The University of Texas Austin, 2515 Speedway, Stop C1400, Austin, TX 78712 USA\\
$^{2}$ TAPIR, California Institute of Technology, Pasadena, CA 95616, USA\\
$^{3}$ Department of Physics and Astronomy, University of California, Davis, CA 95616, USA\\
$^{4}$ Department of Astronomy and Theoretical Astrophysics Center, University of California Berkeley, Berkeley, CA 94720\\
$^{5}$ Department of Physics and Astronomy, University of California, Irvine, CA 92697, USA \\
$^{6}$ Center for Astrophysics | Harvard \& Smithsonian, 60 Garden Street, Cambridge, MA 02138, USA\\
$^{7}$ Harvard Society of Fellows, 78 Mount Auburn Street, Cambridge, MA 02138, USA\\
$^{8}$ Department of Astrophysical Sciences, Princeton University, Peyton Hall, Princeton, NJ 08544, USA
}

\date{Accepted XXX. Received YYY; in original form ZZZ}

\pubyear{2022}

\begin{document}
\label{firstpage}
\pagerange{\pageref{firstpage}--\pageref{lastpage}}
\maketitle

\begin{abstract}
We perform cosmological hydrodynamical simulations to study the formation of proto-globular cluster candidates in progenitors of present-day dwarf galaxies $(M_{\rm vir} \approx 10^{10}\, {\rm M}_\odot$ at $z=0$) as part of the ``Feedback in Realistic Environment'' (FIRE) project. Compact ($r_{1/2}<30$ pc), relatively massive ($0.5 \times 10^5 \lesssim M_{\star}/\msun \lesssim 5\times10^5$), self-bound stellar clusters form at $11\gtrsim z \gtrsim 5$ in progenitors with $M_{\rm vir} \approx 10^9\,\msun$. Cluster formation is triggered when at least $10^7\,\msun$ of dense, turbulent gas reaches $\Sigma_{\rm gas} \approx 10^4\, {\rm M}_\odot\, {\rm pc}^{-2}$ as a result of the compressive effects of supernova feedback or from cloud-cloud collisions. The clusters can survive for $2-3\,{\rm Gyr}$; absent numerical effects, they would likely survive substantially longer, perhaps to $z=0$. The longest-lived clusters are those that form at significant distance --- several hundreds of pc --- from their host galaxy. We therefore predict that globular clusters forming in progenitors of present-day dwarf galaxies will be offset from any pre-existing stars within their host dark matter halos as opposed to deeply embedded within a well-defined galaxy. Properties of the nascent clusters are consistent with observations of some of the faintest and most compact high-redshift sources in \textit{Hubble Space Telescope} lensing fields and are at the edge of what will be detectable as point sources in deep imaging of non-lensed fields with the \textit{James Webb Space Telescope}. By contrast, the star clusters' host galaxies will remain undetectable.
\end{abstract}




\section{Introduction}

It has been over two hundred years since William Herschel declared that globular clusters (GCs) ``are generally but little known and are undoubtedly the most interesting objects in the heavens" \citep{herschel1814}, and while GCs have been the subject of intense and detailed study, many aspects of their formation and evolution remain but little known. While GCs are ubiquitous in massive ($L \gtrsim 0.1\,L^{\star}$) galaxies at $z=0$, exactly how and when they typically form are topics of considerable debate. Some observations of both individual GCs and GC systems around galaxies are naturally reproduced if metal-poor GC formation is connected to specific conditions present only in the high-redshift Universe \citep{peebles1968, moore2006}.  On the other hand, the existence of metal-rich GCs and observations of dense, massive star clusters forming in extreme settings such as mergers in the low-redshift Universe point to a connection between GC formation and the high pressure, high surface density tail of the distribution of star-forming gas in galactic disks \citep{ashman1992, elmegreen1997}. 

Broadly speaking, these can be thought of as \textit{pre-galactic} and \textit{galactic} models, respectively. Although both classes of model are capable of reproducing many bulk properties of the GC population at $z=0$, they differ in the typical epoch of cluster formation --- in or near the epoch of reionization, $z \sim 6-10$, for pre-galactic models, and near, but prior to, the peak of the cosmic star formation history at $z \sim 2-3$ for the galactic models. As a result, the abundance and properties of GCs forming in the reionization era have the potential to definitively discriminate between formation models.

It is therefore momentous that we appear to be on the cusp of directly observing the formation of GCs in the high-redshift Universe. Recent results indicate that many high-$z$ ``galaxies" have properties similar to nascent star clusters (or star cluster complexes; \citealt{ishigaki2018,bouwens2021a,bouwens2021b}), and individual systems magnified by gravitational lensing have revealed clear candidates for GC-like objects in formation at $z \sim 3-6$ \citep{vanzella2017, johnson2017, vanzella2019,vanzella2021}. The \textit{James Webb Space Telescope} (JWST) will likely unveil huge numbers of GCs in formation \citep{carlberg2002,renzini2017,mbk2017,mbk2018,pozzetti2019}, thereby providing a wealth of information about the ``how, when, and where" of GC formation. 

Recent progress in modeling the formation of GCs has also been substantial, with a variety of approaches providing frameworks for understanding the formation of GCs both within their host galaxies and in the broader context of galaxy formation theory. Ongoing work on this front includes (1) simulations --- at very high resolution but without a full cosmological context --- of cluster formation within galaxies or molecular cloud complexes  \citep{he2019, li2019a, lahen2020, lee2020, lahen2021, li2021, hislop2022}; (2) numerical or semi-numerical models of GC formation (which track GCs in cosmological context without directly resolving their formation; e.g., \citealt{katz2014, ricotti2016, li2017, renaud2017, pfeffer2018, creasey2019, el-badry2019, carlberg2020, halbesma2020, phipps2020, reina-campos2022}); (3) simulations linking GC formation to specific conditions at in the high-redshift Universe (e.g., \citealt{mandelker2018, madau2020, lake2021}); and (4) empirical models connecting GCs to high-redshift halos (e.g., \citealt{trenti2015, mbk2017, valenzuela2021}). 

Given the inexorable increase in computing power, it is also now possible to directly resolve GC formation in cosmological simulations of galaxy formation \citep{kimm2016,kim2018, Ma2020}. Most of these focus on galaxies that are fairly massive relative to a typical galaxy (i.e., $M^\star(z)$), in large part because the ubiquity of GCs in $L \gtrsim 0.1\,L^{\star}$ galaxies at $z=0$ guarantees the conditions for GC formation are universally met at some time in such objects. The presence of GCs in many nearby \textit{dwarf} ($M_{\star} \lesssim 3\times 10^9\,M_{\odot}$) galaxies may be an important clue to the origin of GCs more generally. 

GCs are present even in galaxies as faint as $M_{\star} \approx 10^5\,M_{\odot}$ (Eridanus II; \citealt{bechtol2015,koposov2015,crnojevic2016, simon2021}), a regime in which standard models of galaxy formation predict that the majority of star formation should have occurred by the end of the reionization era, $z \sim 6$ \citep{bullock2000, benson2002, somerville2002, ricotti2005}. The fraction of stars contained in GCs in low-mass ($M_\star \lesssim 10^{7}\,\msun$) galaxies is $1-10\%$ \citep{georgiev2010, hudson2014, larsen2017}, which is much higher than in more massive systems and indicates that cluster-related star formation plays an important role in the growth of these systems; this is especially true at early times, as the clusters can contain ${\sim}25\%$ of the metal-poor ($[{\rm Fe/H}]<-2$) stars in dwarf galaxies \citep{larsen2014}. However, the fraction of galaxies hosting at least one GC drops off strongly towards low stellar masses \citep{georgiev2010, burkert2020, eadie2021}, meaning that conditions for GC formation are met in only a subset of dwarf galaxies; an understanding of which dwarf galaxy progenitors achieve the necessary conditions for GC formation may prove essential for understanding GC formation more broadly. 

In this paper, we perform a series of cosmological zoom simulations to study the formation of bound stellar clusters in the progenitors of present-day dwarf galaxies ($M_{\rm halo}(z=0)\approx 10^{10}\,M_{\odot}$). We refer to the star clusters of interest as proto-globular cluster candidates (GCCs). In these proof-of-principle simulations, we focus on the conditions that lead to the formation of dense star clusters at high redshift in such galaxies and explore when this cluster formation occurs, the masses and lifetimes of these clusters, and the properties of the clusters relative to their host galaxies (e.g., the mass of the cluster relative to the host galaxy and the location of formation of the clusters with respect to the size of the galaxies). This paper is organized as follows: in Section \ref{sec:sim}, we discuss our simulation setup. Section~\ref{sec:clusters} describes the formation and properties of the clusters and connections to their larger-scale environments. Section~\ref{sec:discuss} discusses implications for the detectablity of clusters in situ in the high-redshift Universe and for GC formation models as well as the sensitivity of our results to variations in the treatment of galaxy formation physics and numerics in the simulations. In Section~\ref{sec:conclusion}, we present our conclusions. Unless otherwise noted, all lengths quoted in this paper are physical, not comoving.

\begin{table*}
\begin{adjustbox}{width=\textwidth,center}
    \centering
    \setlength\tabcolsep{1.0pt}
    \renewcommand{\arraystretch}{1.15}
    \begin{tabular*}{\textwidth}{l @{\extracolsep{\fill}} | cccccccc|cccccc}
        \hline
        & \multicolumn{8}{c|}{{\sc Cluster(s)}}  & \multicolumn{6}{c}{ {\sc Host dark matter halo \& galaxy} (at $z_{\rm f})$} \\

        \hline
        Halo & $z_{\rm f}$ & $r_{1/2}$ & $M_{\star,\rm cl}$ & d & $t_{50}$&$\Delta t_{\rm form}$ &[Fe/H]& $\sigma_{\rm [Fe/H]}~$ &$M_{\rm vir}$ &  $r_{\rm vir}$ & $V_{\rm m}$  & $M_\star$& $r_{\rm 1/2,h}$ &${\rm [Fe/H]_h}$\\
        & & (pc) &  ($\text{M}_\odot$) & (pc) & (Myr)& (Myr) & && ($\text{M}_\odot$) &  (kpc)&  (km/s) & ($\text{M}_\odot$) & (pc) & \\
        \hline
        m10b & --- & --- & --- & --- & --- & --- & --- & --- & --- & --- & --- & --- & --- & --- \\
        \hline
        \hline
        m10i & 11.0 & 26 & $2.3\times10^5$ & 325 & 197 & 6.9 & -2.5 & 0.20 & $5.1\times10^8$ & 2.2 & 34 & $3.9\times10^4$ & 124 & -3.1\\
        m10i$^\dagger$ & 4.9 & 27 & $1.6\times10^5$ & 310 & 270 & 9.6 & -2.7 & 0.17 & $1.3\times10^9$ & 5.8 & 33 & $1.6\times10^5$ & 175 & -3.0\\
        \hline
        m10l$^*$ & 8.1 & 17 & $2.2\times10^5$ & 802 & 1800 & 7.4 & -2.9 & 0.19 & $7.4\times10^8$ & 3.2 & 34 & $2.1\times10^5$ & 171 & -2.9\\
        \hline
         m10j  & 9.0 & 18 & $1.3\times10^5$  & 1280 & 1620 & 9.2 & -2.3 & 0.14 & $4.4\times10^8$  & 2.4 &  29  & $4.9\times10^4$ & 85 & -3.4\\
         \hline
         m10k  & 4.5 & 19 &  $3.6\times10^5$  & 60 & 940 & 7.2 & -2.8 & 0.18 & $1.2\times10^9$ & 6.4 & 29 & $1.6\times10^5$ & 169 & -2.8\\
         \hline
         m10m$^*$~ & 6.0 & 15 & $4.5\times10^5$ & 628 & 2500 & 7.3 & -2.8 & 0.16 & $1.6\times10^9$ & 5.4 & 39 & $1.6\times10^5$ & 320 & -3.0\\
         \hline
         m10h & 5.7 & 6 & $5.3\times10^4$ & 614 & 94 & 4.2 & -2.7 & 0.17 & $1.1\times10^9$ & 5.0 & 33 & $1.1\times10^5$ & 370 & -2.7\\
         \hline
    \end{tabular*}
     \end{adjustbox}
    \caption{Properties of the stellar clusters and their host galaxies, all computed at the epoch of cluster formation. Columns specify:
     (1) $z_{\rm f}$: redshift the cluster has formed; (2) $r_{1/2}$: 3D stellar half-mass radius of the cluster at the formation time; (3) $M_{\rm cl}$: mass of each cluster at the formation epoch; (4) $d$: the distance from its host; (5) $t_{50}$: the cosmological time since the formation epoch for each cluster that has lost 50\% of its mass; (6) $\Delta t_{\rm form}$: Formation time window in which all the cluster members have formed; (7) [Fe/H]: the iron abundance of the cluster; (8) $\sigma_{\rm [Fe/H]}$: the spread in the iron abundance; (9) $M_{\rm vir}$: host DM halo virial mass; (10) $r_{\rm vir}$: host's virial radius; (11) $V_{\rm m}$: host's maximum circular velocity; (12) $M_\star$: stellar mass (\textit{excluding} the stellar mass contained in the nascent clusters) within $0.1\,r_{\rm vir}$ of the host halo's center; (13) $r_{\rm 1/2,h}$: host's 3D stellar half-mass radius; (14) ${\rm [Fe/H]_h}$: iron abundance for all stars associated to the host. The m10i simulations forms two clusters with  $M_{\rm cl}>10^5\, {\rm M}_\odot$ at different redshifts, with the later-forming cluster (denoted m10i$^\dagger$ in the table) hosted in a separate halo that is not part of the merger tree of the main halo (but is still within the high-resolution region of the simulation). The clusters denoted with an asterisk, m10l and m10m, each form two coeval clusters that form separately out of a single massive GMC within the main galaxy and shortly thereafter merge together to form one massive, long-lived cluster. One halo, m10b, does not form any bound clusters in excess of $M_\star=5\times 10^4\,\msun$.}
    \label{tab:data}
\end{table*}
\section{Simulations}\label{sec:sim}
Our simulation suite is part of the ``Feedback In Realistic Environment" project \citep[FIRE,][]{hopkins2014,hopkins2018}\footnote{\url{https://fire.northwestern.edu}}. We select seven realizations of halos with virial masses of ${\sim}10^{10}\, \text{M}_\odot$ at $z=0$ --- hosts of present-day dwarf galaxies --- from \citet{fitts2017} and re-simulate them with an updated version of the {\sc GIZMO}\footnote{\url{http://www.tapir.caltech.edu/~phopkins/Site/GIZMO.html}} code \citep{hopkins2014} and FIRE-2 galaxy formation prescriptions \citep{hopkins2018}. These simulated halos comprise the 6 galaxies with the highest $z=0$ stellar mass in the \citet{fitts2017} suite (m10h, m10i, m10j, m10k, m10l, m10m) along with the one of the lowest $M_\star(z=0)$ galaxies (m10b). The gravity solver in {\sc GIZMO} is a descendant of {\sc GADGET3} (first described in \citealt{springel2008}) and the hydrodynamical equations are treated via the mesh-free finite mass (MFM) Lagrangian Godunov method, which provides adaptive spatial resolution while maintaining conservation of mass, energy, and momentum. We adopt a flat $\Lambda$CDM cosmology with $h=0.71$, $\Omega_{\rm m}=0.266=1-\Omega_{\Lambda}$, and $\Omega_{\rm b}=0.0449$, all consistent with  7-year data from {\it WMAP} \citep{komatsu2011}. These differ slightly from the latest constraints based on full-mission \textit{Planck} observations \citep{planck2018}, but these differences are unimportant for the topics considered here. 

The initial conditions (ICs)\footnote{All ICs can be found at \url{http://www.tapir.caltech.edu/~phopkins/publicICs}.} are generated using the zoom technique \citep{katz1993, onorbe2014} at $z=125$, embedded within periodic cosmological boxes of $L=25\, {\rm Mpc}/h$, and are computed via the code {\sc MUSIC} \citep{hahn2011}. The ICs have particle masses of $m_{\rm gas}=500\, {\rm M}_\odot$ and $m_{\rm dm}=2500\, {\rm M}_\odot$, and we perform the simulations with the Plummer-equivalent adaptive gravitational softening lengths of $\epsilon_{\rm gas}=2$ pc, $\epsilon_\star=3.7$ pc, and $\epsilon_{\rm dm}=35$ pc (physical). For each simulation, we output snapshots every 10 Myr --- comparable to the expected formation period of massive star clusters \citep{bastian2014,hollyhead2015}  --- over the period $z=15-4$ (yielding 134 snapshots in this redshift range) to enable detailed study of the formation of any bound stellar clusters within this redshift range. For $z<4$, we output snapshots every 250 Myr, resulting in 180 snapshots in total. We identify and characterize haloes and subhaloes using a modified version of the code {\sc ROCKSTAR} \citep{behroozi2013,wetzel2020b,wetzel2020}.

Our simulations explicitly resolve the multi-phase interstellar medium (ISM), with heating and cooling of gas modelled in the temperature range of $T=10-10^{10}$ K. Star formation happens in self-shielding, self-gravitating, Jeans-unstable gas clouds above a density threshold of $n_{\rm thresh}=1000\, {\rm cm}^{-3}$. 
Our stellar feedback model includes SNe Ia \& II, multi-wavelength photo-heating, cosmic ray heating, stellar winds, and radiation pressure, all adopted from {\sc STARBURST99} stellar evolutionary model \citep{leitherer1999} assuming a Kroupa initial mass function \citep{kroupa2002}. The ultraviolet background radiation model is an updated version of the original model presented in \citet{fg2009}\footnote{\url{http://galaxies.northwestern.edu/uvb/}.} and completes the reionization around $z=6$. Relative to earlier FIRE simulations, including the \citet{fitts2017} suite, our fiducial FIRE-2 model includes metal-diffusion physics, a slightly updated ultraviolet background radiation model, and a correction to the cosmic ray heating source term that avoids spurious heating at very early times in the simulations. We also adopt an updated stellar mass-loss algorithm (as we discuss further in section \ref{subsec:sensitivity}) that differs from the default FIRE-2 implementation outlined in \citealt{hopkins2018} but follows the default treatment in FIRE-3 \citep{hopkins2022}. In a second set of simulations, 
we use the default FIRE-2 algorithm for the stellar mass-loss processes (as described in \citealt{hopkins2018} and in section \ref{subsec:sensitivity} below) or vary the density threshold from our standard choice of $n_{\rm thresh}=1000\, {\rm cm}^{-3}$, keeping all other fiducial FIRE-2 physics unchanged, to study the impact on star formation activity and any potential difference in efficiency of galaxies in forming GCCs. 

We identify star clusters using the {\sc Phinder} algorithm first described in \citet{grudic2018b}. {\sc Phinder} searches for local minima of the stellar gravitational potential and identifies bound particles within each group. We only keep clusters with at least 32 bound members, resulting in a minimum cluster mass of $M_{\rm cl}\approx1.6\times 10^4\, {\rm M}_\odot$, in our cluster catalogs. However, even this conservative choice typically results in bound objects with very large half-mass radii $R_{1/2}>100$ pc. These objects get destroyed very quickly in the tidal forces of the interstellar medium, meaning they are not good candidates for long-lived and self-bound star clusters. In our analysis, we only consider bound clusters with initial stellar mass of $M_{\rm cl}=100\,m_\star\approx5\times10^4\, {\rm M}_\odot$ and 3D half-mass radii of $r_{1/2}<50$ pc. This choice results in the selection of long-lived, bound stellar clusters, as we show below. Given that our chosen force softenings for baryonic particles are comparable to half-light radii of the most compact GCs, we do not expect to resolve the true internal structure of our bound clusters in this proof-of-concept work. Since our simulations employ softened rather than direct gravitational force calculations, the internal dynamics of the clusters could not be accurately tracked over cosmological times even had they formed with the smaller sizes of present-day GCs. 

Our analysis results in several GCCs across our simulation suite. Most form within the most massive halo in the simulation volume; the exceptions are the m10k cluster and the second cluster in the m10i realization. Though many lower-mass cluster candidates are identified using this procedure, we focus only on clusters with $M_{\star}>5\times 10^{4}\,\msun$ ($>100$ star particles at birth) for the remainder of this paper. Of the seven halos from the \citet{fitts2017} suite that we have resimulated, six contain such a star cluster, and each of these halos forms only a single cluster with $M_{\star} > 10^5\,\msun$. The lone exception, halo m10b, has a central galaxy with a much lower stellar mass at $z=0$ based on the \citet{fitts2017} suite, a factor of 15 less than the other halos simulated here, providing a tentative indication of a connection between total stellar mass formed and the presence of massive star clusters at fixed $z=0$ halo mass; we return to this point in Section~\ref{sec:discuss}. In the following sections, we discuss the formation of the clusters, their properties, and connections with their host galaxies and dark matter halos.

\section{Formation and properties of stellar clusters} 
\label{sec:clusters}
The basic properties of each cluster are listed in the first several columns of Table~\ref{tab:data}. The clusters form between redshift 4.5 and 11.0 and have stellar masses that range from $5\times 10^{4}$ to $5\times 10^5\,\msun$ at formation. The 3D half-mass sizes of the clusters are all smaller than 30 pc; as noted in Sec.~\ref{sec:sim}, the sizes we find here should be considered upper limits, as our adopted force softening for baryonic particles ($2-4\,{\rm pc}$) precludes the formation of significantly denser systems. The clusters typically form with very low metallicities ($[{\rm Fe/H}] \approx -2.5~{\rm to}\,-3$), with dispersion in the iron metallicty of $0.15-0.2$ dex. These metallicities are comparable to or slightly lower than those of the most metal-poor clusters observed in low-mass galaxies \citep{beasley2019}, which we discuss further in Sec.~\ref{subsec:previous}; the spread in metallicity is somewhat larger than observed in MW GCs ($0.045$~dex; \citealt{carretta2009, bailin2019}). The iron spread in the simulated clusters is inherited from the progenitor gas clouds as opposed to resulting from self-enrichment during the star formation process. We define the formation time $t_{\rm form}$ of the cluster as the time when the first ten members of the cluster form. Our results are insensitive to this precise definition, as the full duration of star formation is less than 10 Myr in all of the clusters studied here.

\begin{figure*}
    \centering
    \includegraphics[width=\textwidth]{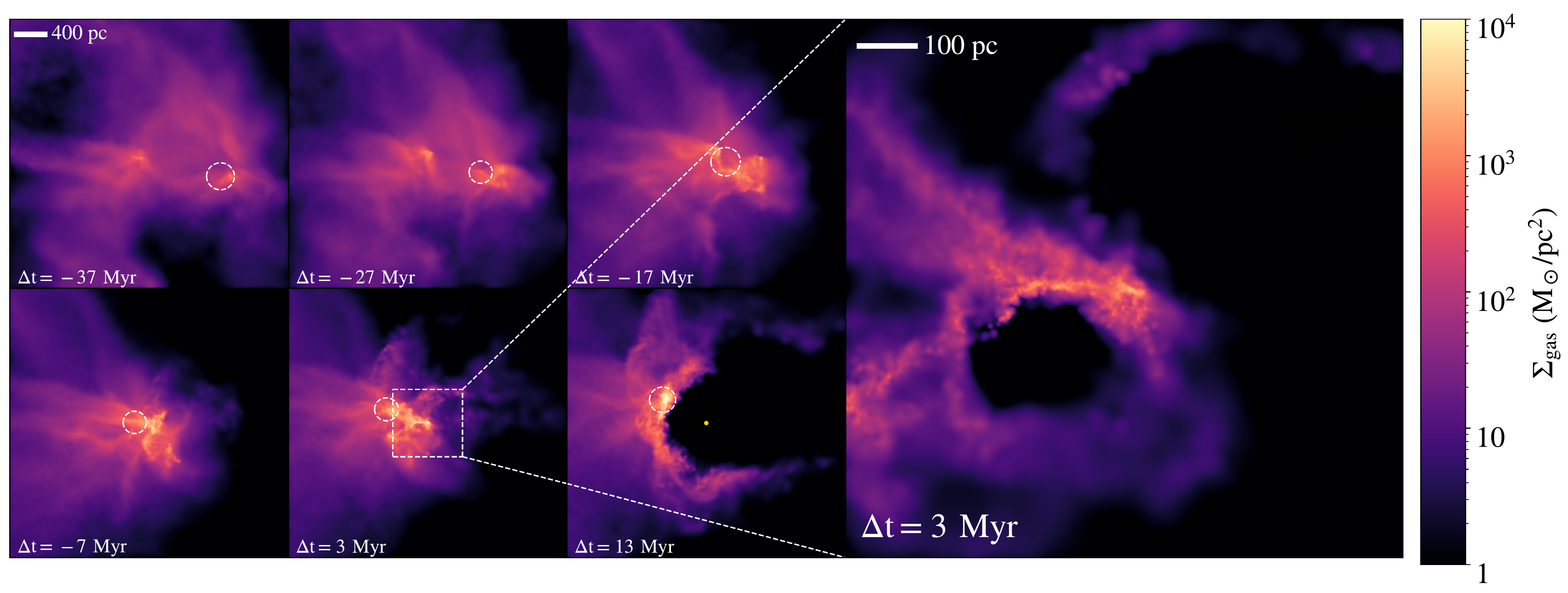}\\
    \includegraphics[width=\textwidth]{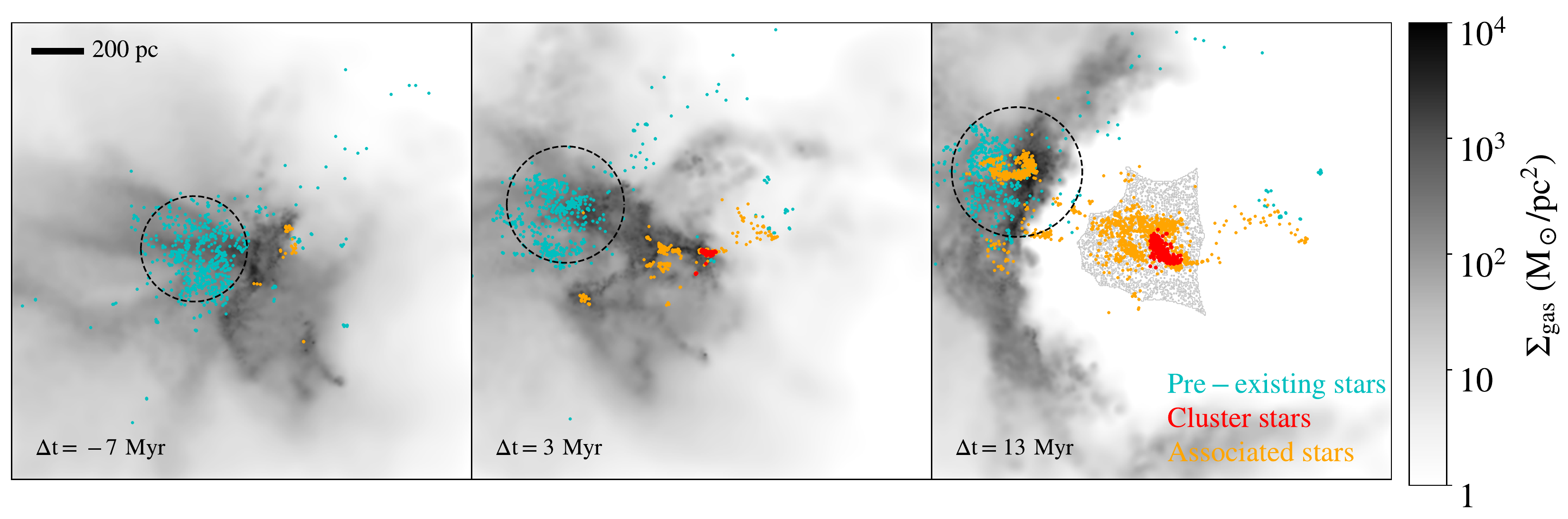}
    \caption{{\bf Top}: Time evolution of the gas surface density of the progenitor gas cloud that gives birth to a bound stellar cluster in m10l galaxy at $z=8.1$. Each smaller sub-panel is $(4\,{\rm kpc})^2$ with a depth of $1\,\text{kpc}$, centered on the center of mass of the gas progenitors that eventually give birth to the cluster members. The time relative to the cluster formation epoch is noted in each frame and the pre-existing galaxy is indicated with a white circle with size equal to its stellar half-mass radius. The larger sub-panel on the right-hand side, spanning $(1\, {\rm kpc})^2$ with a depth of $0.5\, \text{kpc}$, zooms into the region where the cluster forms just prior to its birth. At this epoch, a substantial amount of gas, $\approx 10^7\,\msun$, reaches a surface density of $\Sigma_{\rm gas} \sim 10^4\, \msun \, {\rm pc}^{-2}$, a necessary condition for cluster formation, as a result of a cloud-cloud collision. The cluster itself is shown at $\Delta t=13\,{\rm Myr}$ as a gold circle with size equal to the cluster half-mass radius (20 pc).
    {\bf Bottom}: $(2\,{\rm kpc})^2$ gas surface density maps, with depth of $1\,{\rm kpc}$, at the last three times shown in the upper panels. Pre-existing stars are shown in cyan, stars that are part of the bound cluster that forms are shown in red, and stars forming from $\Delta t = -17~{\rm Myr}$ to the snapshot in question that are \textit{not} bound to the cluster are shown in orange. The cloud-cloud collision results in a very high surface density \textit{outside} of the pre-existing galaxy (whose stellar half-mass radius is shown in black in each panel). This gas is further compressed by supernovae resulting from stars born in the colliding gas (orange in left panel), leading to the formation of the stellar cluster. Many stars that are not formally identified as cluster members are formed at the same time in the immediate vicinity of the cluster as well (right panel), and feedback from the cluster formation launches a compressive shock that causes star formation within the pre-existing galaxy.
    }
    \label{fig:m10l}
\end{figure*}

\begin{figure*}
    \centering
    \includegraphics[width=\textwidth]{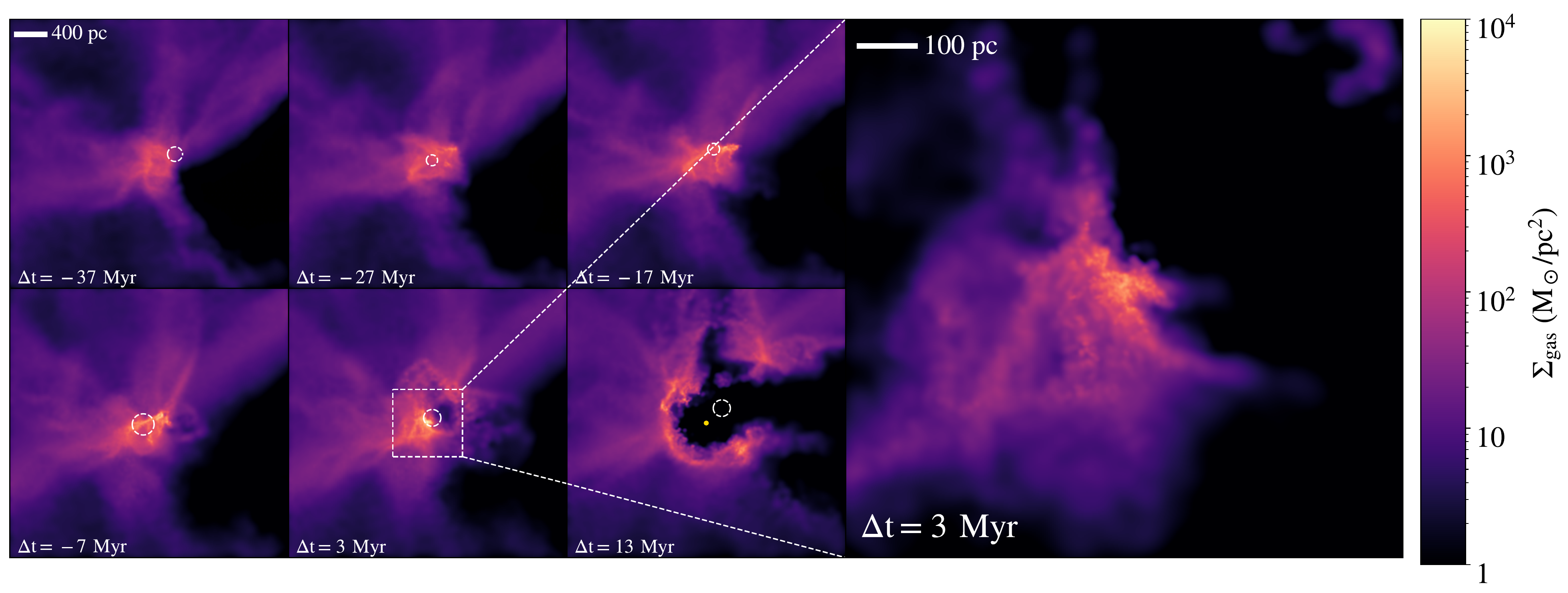}
    \includegraphics[width=\textwidth]{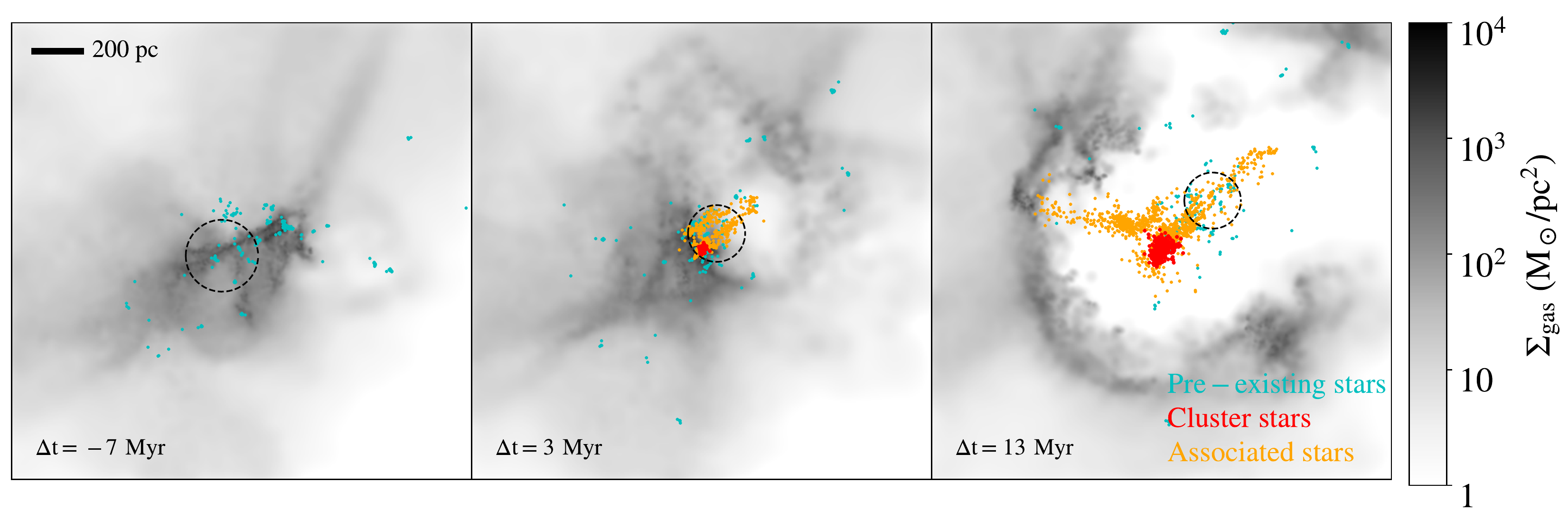}
    \caption{Analogous to Fig.~\ref{fig:m10l} but for the bound stellar cluster formed in m10i galaxy at $z=10.8$. Unlike the cluster in m10l, there is no cloud-cloud collision inducing the formation of the cluster. In this case, it is filamentary gas accretion, coupled with compression due to supernova feedback, that pushes gas to $\Sigma_{\rm gas}=10^4\, {\rm M}_\odot\, {\rm pc}^{-2}$. In this case, the cluster formation happens at such an early epoch that there is barely even a ``host galaxy" to speak of, and the stellar mass formed as a result of the cluster formation far exceeds the pre-existing stellar content of the halo.
    }
    \label{fig:m10i}
\end{figure*}

\subsection{Cluster formation mechanisms}\label{sec:formation-mechanism}
Fig. \ref{fig:m10l} shows the evolution of the progenitor gas cloud in the m10l halo prior the formation of a cluster at $z=8.1$. The top portion contains six panels showing the time evolution of the gas surface density $\Sigma_{\rm gas}$ in the cluster's environment prior to and immediately following the cluster formation epoch (times relative to the formation of the cluster, $\Delta t \equiv t-t_{\rm form}$, are given in each panel). Each panel spans $(4\, \text{kpc})^2$ with a depth of $1~{\rm kpc}$ and is centered on the center of mass of the gas progenitors that eventually give birth to the cluster members; $\Sigma_{\rm gas}$ is computed in cells with areas of $(4\,{\rm pc})^2$ and depths of $1~{\rm kpc}$. The larger subplot on the right is a zoomed-in view of the gas cloud in the midst of cluster formation (with dimension of $(1\, \text{kpc})^2 \times 0.5~{\rm kpc}$). The central galaxy of the halo, with center coincident with the center of the halo, is indicated in each subpanel by a white circle, with radius equal to the galaxy's stellar half-mass radius at that time. The newly-formed cluster is shown as a gold circle in the final panel, with size equal to its half-mas radius; it is significantly offset from the galaxy at this epoch.

The panels show two dense patches of gas approaching one another with halo-centric speeds comparable to the $v \approx 30\,{\rm km/s}$ virial velocity of the halo (which is also comparable to the turbulent velocity dispersion of the gas) and colliding, leading to a region of very high pressure and density by $\Delta t=-7 \,{\rm Myr}$. These are the requisite conditions for cluster formation. It is evident that feedback has cleared gas from the cluster's immediate vicinity by $\Delta t=13\,{\rm Myr}$, leading to a compressive super-bubble that stimulates high gas densities hundreds of pc from the cluster. 

In the bottom panels of Fig. \ref{fig:m10l}, we show gray-scale gas surface density maps, each with dimension $(2 \,{\rm kpc})^2$ with a depth of 1~kpc, at the three snapshots closest to the epoch of cluster formation ($\Delta t=-7, \,3$, and 13~Myr). The stars formed before $\Delta t=-7\,{\rm Myr}$ are shown in cyan, while the stars formed during the epoch of cluster formation are shown in red (for particles identified by {\sc Phinder} as cluster members) and gold (for stars identified as non-cluster members). The half-light radius of the pre-existing galaxy is shown as a black circle in each panel. The cloud-cloud collision results in very high surface densities near, but outside of, the pre-existing galaxy at $-7\,{\rm Myr}$. Feedback from a handful of stars formed in this gas causes further compression, initiating the formation of two coeval sub-clusters in very close proximity and a smattering of other stars. By $\Delta t=13\,{\rm Myr}$, the cluster stars have formed, along with a significant population of nearby stars that are not formally associated with the cluster but nonetheless form in its immediate environment, at distances of $<200\,{\rm pc}$. Feedback from these combined populations clears the gas from the cluster region and launches a bubble, which collides with the pre-existing galaxy and initiates further star formation there as well; we discuss this extra-cluster star formation further in Section~\ref{subsec:galaxy_props}. Finally, the two adjacent sub-clusters merge within ${\sim 40}~{\rm Myr}$ of their formation, with each sub-cluster contributing roughly half of the final stellar mass in the merged cluster; the cluster in halo m10m follows a very similar formation channel.

Figure~\ref{fig:m10i} is analogous to Figure~\ref{fig:m10l} but shows the gas surface density evolution immediately preceding the formation of a cluster in the m10i halo. In this case, there is no cloud-cloud collision; rather, filamentary gas accretion leads to high surface density near the center of the halo. This high surface density initially promotes clustered star formation but not the formation of a cluster ($\Delta t=-(17-7)$~Myr). However, feedback from the subsequent SN explosions resulting from that star formation drives a bubble that compresses the already dense ISM even further, resulting in the formation of a cluster. Within 10~Myr, feedback from the lives and deaths of massive stars in the cluster has expelled all of the gas and star formation ceases.

\begin{figure*}
    \centering
    \includegraphics[width=\textwidth]{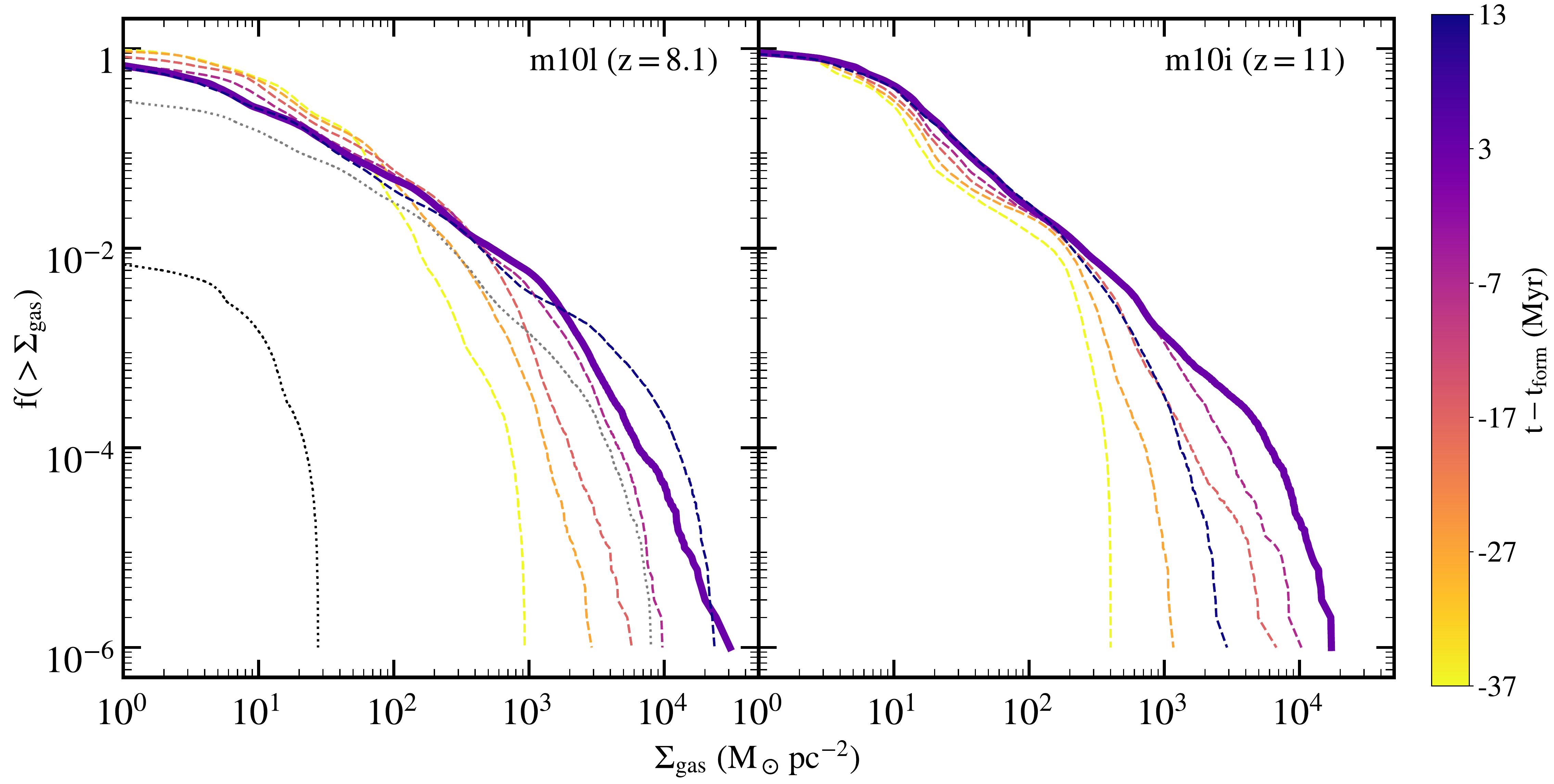}\\
    \includegraphics[width=\textwidth]{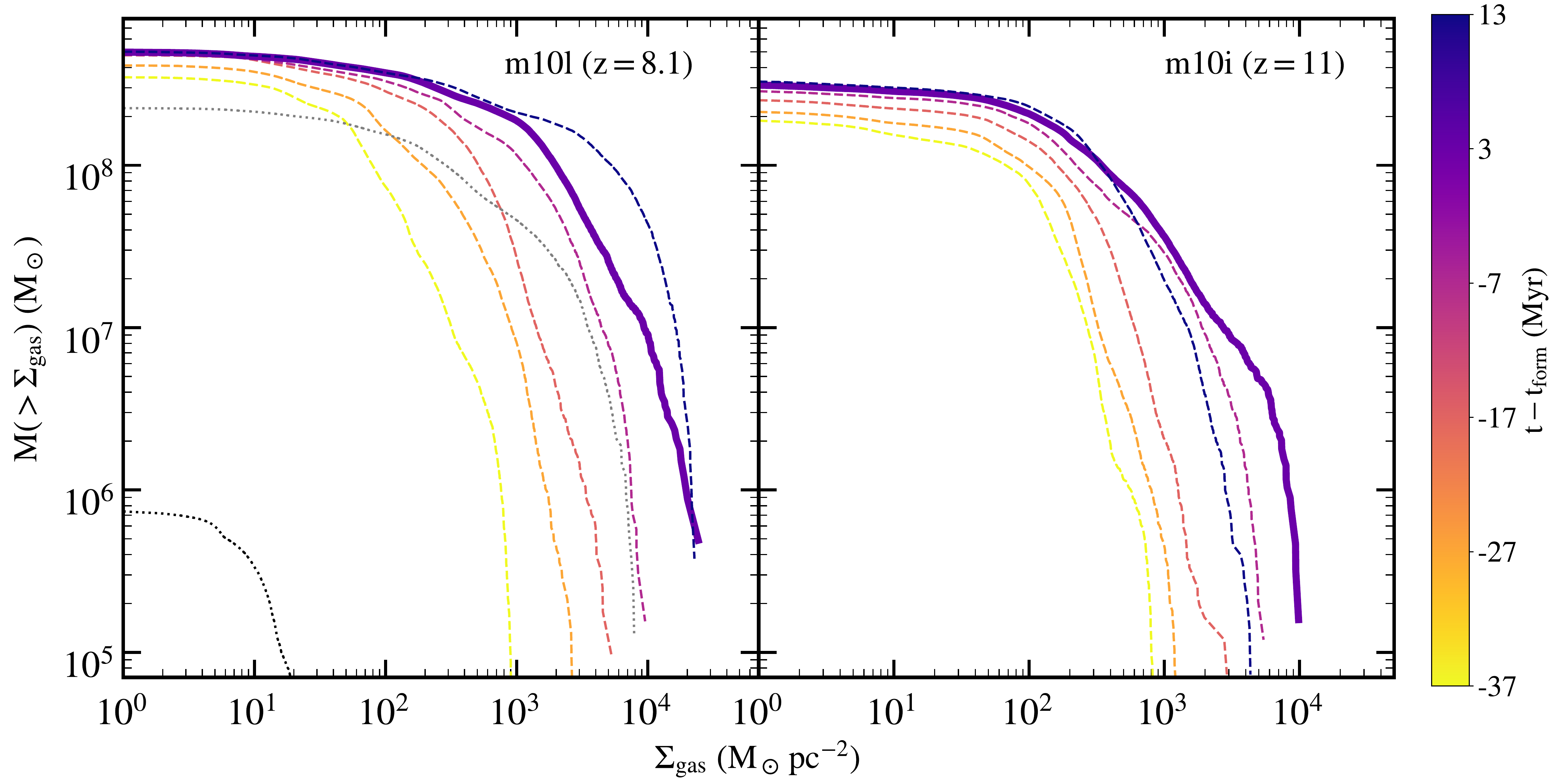}
    \caption{\textbf{Upper panels}: The cumulative distribution of gas cells ($4\times 4\times 1000\,{\rm pc}$ each) exceeding a surface density $\Sigma_{\rm gas}$ in 10 Myr increments immediately prior to and during the cluster formation epoch in cluster m10l (left) and m10i (right); line colors indicate the time relative to formation as shown in the color bars on the right, with the thick solid line corresponding to $\Delta t=3~{\rm Myr}$, which is the epoch closest to $t_{\rm form}$. \textbf{Lower panels}: The cumulative distribution of gas mass in the same cells exceeding $\Sigma_{\rm gas}$ for the same snapshots. For halo m10l, we also show both quantities at $\Delta t=23$ Myr (gray dotted line) and 33 Myr (black dotted line). Cluster formation occurs when $\gtrsim 10^7\,\msun$ of gas exceeds $10^4\, {\rm M}_\odot\, {\rm pc}^{-2}$, a result that holds true for all clusters in our simulation suite. The gas surface density drops precipitously after the cluster forms in all cases. Generally, the drop occurs very shortly after cluster formation ($\Delta t = 10\,{\rm Myr}$. In m10l, an even steeper drop occurs but is delayed by 20~Myr because of the additional star formation triggered by feedback from the cluster's formation (see Fig.~\ref{fig:m10l}). However, in all cases, the cluster formation site itself is cleared of gas within 10~Myr of cluster formation.}
    \label{fig:surface_density}
\end{figure*}

Figures~\ref{fig:m10l}~and~\ref{fig:m10i} make it clear that there are multiple pathways to get to high densities and pressures that are conducive to cluster formation. To better quantify what constitutes ``high density" and how this is linked to the formation of stellar clusters, we compute $\Sigma_{\rm gas}$ in cells of $4\times 4\times 1000\,{\rm pc}$, centered on center-of-mass of gas particles that give birth to the cluster, and plot the cumulative distribution of this quantity in 10 Myr intervals over a 50 Myr window spanning cluster formation, $-37 \leq \Delta t/{\rm Myr} \leq 13$, in Figure~\ref{fig:surface_density}. The colors of the lines indicate the time relative to $t_{\rm form}$ via the colorbar at the right; the thick solid line corresponds to the snapshot closest to cluster formation, $\Delta t=3~{\rm Myr}$. In both cases highlighted here, the maximum surface density is initially a few hundred $\msun\,{\rm pc}^{-2}$ and quickly rises until ${\sim}0.01\%$ of the cells exceed $\Sigma_{\rm gas} \approx 10^{4}\,{\rm pc}^{-2}$. The bottom panels show the related quantity $M_{\rm gas}(>\Sigma_{\rm gas})$ for the same cells and indicate that ${\sim} 10^{7}\,\msun$ of gas is contained in the cells exceeding $\Sigma_{\rm gas} \approx 10^{4}\,\msun\,{\rm pc}^{-2}$, meaning that the region containing this high density has a surface area of order $(20~{\rm pc})^2$. At this point, the clusters form rapidly: in each case studied in this paper, all star formation in the cluster takes place in less than 10~Myr.

The resulting feedback immediately removes all gas from newly-formed cluster. Accompanying the gas removal from the cluster is a precipitous drop in both $f(>\Sigma_{\rm gas})$ and $M(>\Sigma_{\rm gas})$ at high gas surface densities ($\Sigma_{\rm gas} \gtrsim 10^{3}\,\msun\,{\rm pc}^{-2}$) in m10i. Halo m10l does \textit{not} see an immediate drop (and in fact sees a temporary increase) because of the star formation that is triggered within the galaxy by the feedback that clears the gas from the nascent cluster (see Fig.~\ref{fig:m10l}). However, m10l sees the same drop starting ${\sim }20$~Myr after cluster formation (gray dotted curves in Figure~\ref{fig:surface_density}), and after another 10 Myr (black dotted curves), feedback has removed most of the gas from the region considered here. We conclude that stellar feedback from cluster formation is very effective at clearing gas from the cluster formation site on the cluster formation time scale of $\lesssim 10~{\rm Myr}$, in agreement with observations \citep{bastian2014,hollyhead2015,krumholz2019}.

In the two cases highlighted here, and indeed in all the identified clusters, cluster formation occurs once giant molecular clouds reach surface densities exceeding $\Sigma_{\rm thresh}=10^{4}\,\msun\,{\rm pc}^{-2}$ (see also \citealt{murray2010, colin2013, geen2017, elmegreen2018, grudic2018, kim2018, krumholz2019}). This threshold surface density, which is a factor of ${\sim}100$ higher than is typical of giant molecular clouds (GMCs) in the Milky Way \citep{bolatto2008}, is equivalent to a gravitational force per unit area (i.e., pressure) of $P_{\rm thresh} \equiv G\,\Sigma_{\rm thresh}^2=2 \times 10^9 \,k_{\rm B}\,K \,{\rm cm}^{-3}$. It is also what would be achieved by gas patches with density $n\approx 10^{3}\,{\rm cm}^{-3}$ colliding at a relative velocity of $v_{\rm rel} \approx 120\,{\rm km/s}$ (assuming $P=\rho\,v_{\rm rel}^2$), which is relevant for m10l (see Fig.~\ref{fig:m10l}): gas clumps on ballistic trajectories in a $V_{\rm vir} \approx 30\,\kms$ halo colliding near the halo's center will achieve a gravitational pressure that is close to $P_{\rm thresh}$, and any additional force from SN feedback near the collision will likely be sufficient to achieve $P_{\rm thresh}$. 

The gas surface density can occasionally exceed $\Sigma_{\rm thresh}$ at times that are not linked to the formation of the massive and bound stellar clusters studied in this work. In all of those times, we find lower-mass ($M_\star < 5\times 10^4\,\msun$) clusters forming, and we also find that the amount of mass above $\Sigma_{\rm thresh}$ is commensurately lower: the mass of the cluster(s) that form is roughly a constant fraction of the high surface density gas, $M_{\star} \approx 0.01\,M_{\rm gas}(>\Sigma_{\rm thresh})$. We defer a more thorough investigation of the connection between gas surface density and cluster formation efficiency (and star formation rate) to a future paper.

\subsection{Cluster evolution after formation}
\label{sec:tidal-evolution}
While the formation of star clusters is an important topic in and of itself, the subsequent survival of those clusters is a crucial consideration when attempting to connect compact, bound stellar systems at high redshifts to GCs at $z=0$. 
We study the evolution of the star clusters formed at high redshift in our simulations by tracking clusters across simulation snapshots. To do so, we run {\sc Phinder} separately at each snapshot and then use the IDs of the each cluster at birth (the highest-redshift snapshot at which it is identified) to find its descendant at each later time. The bound clusters formed in the simulations lose mass with time owing to both tidal interactions with their environment and numerical effects. We quantify cluster lifetimes via $t_{50}$, the time relative to its birth at which a cluster has lost 50\% of its original stellar mass; the clusters typically disrupt completely soon after $t_{50}$. We find a range of $t_{50}$ values, from ${\sim}100$~Myr to 2.5~Gyr (see Table~\ref{tab:data}). The longest-lived clusters are those that form at the greatest distances from the center of their host halos, indicating the importance of tides --- coupled with numerical effects --- in disrupting the clusters in our simulations. Notably, only one cluster forms within its host galaxy's half-mass radius and that some form at $>\,4\,r_{\rm 1/2,h}$, effectively completely outside of the galaxy. 

\begin{figure}
    \centering
    \includegraphics[width=\columnwidth]{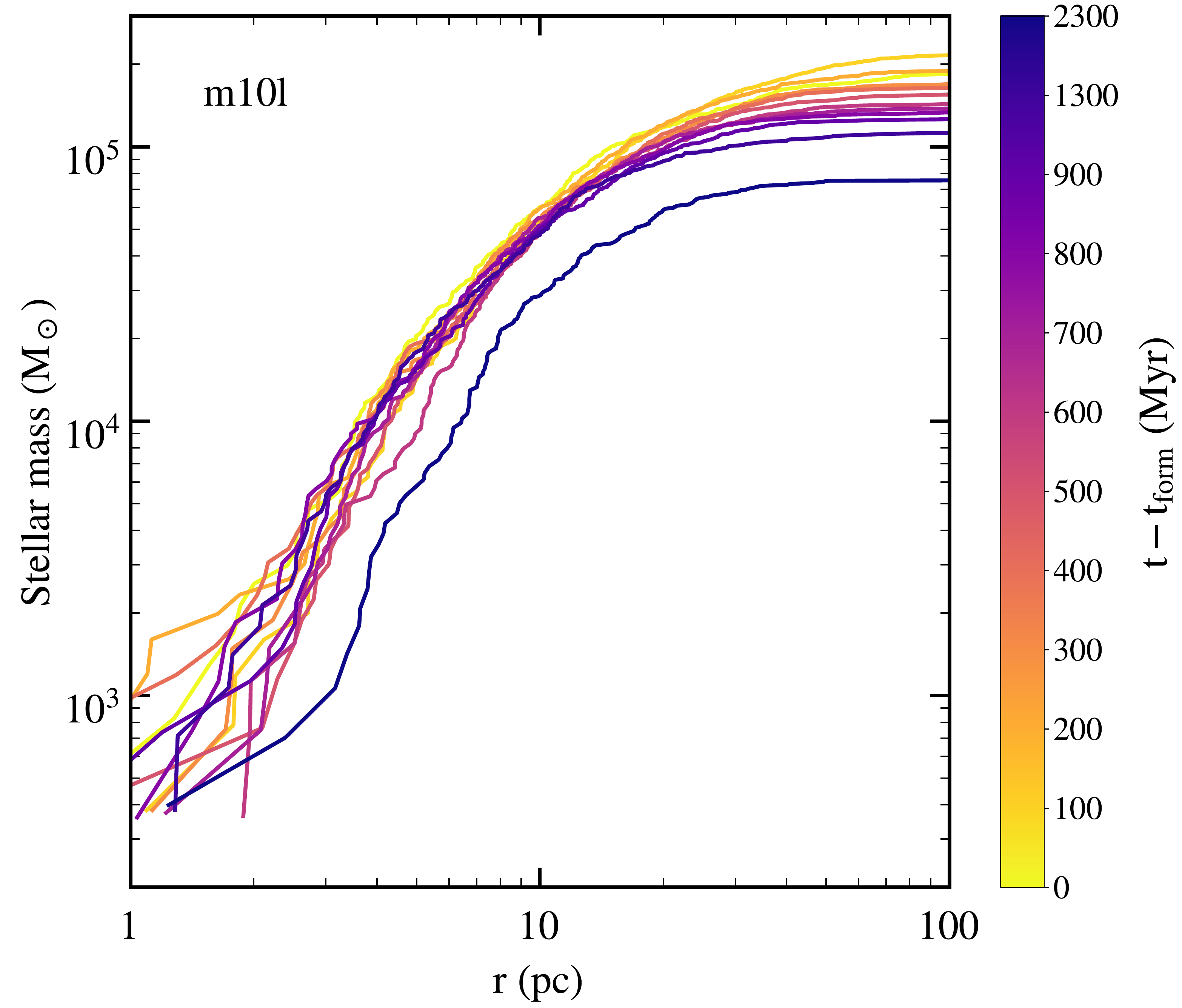}
    \caption{The time evolution of the stellar mass profile of the cluster that forms in the m10l halo. The inner profile is very stable over the ${\sim}1.5\,{\rm Gyr}$, a period over which the cluster completes nearly 15 orbits about its host galaxy. A combination of numerical and tidal effects lead to its dissolution after nearly 2.5~Gyr.}
    \label{fig:mass_profile}
\end{figure}

In Fig. \ref{fig:mass_profile}, we show the time evolution of the stellar mass profile of the (merged) m10l cluster over $2.5\,{\rm Gyr}$ after its formation. For the first ${\sim} 1.3$~Gyr, corresponding to over 100 crossing times, the cluster's inner mass profile is remarkably stable, while secular mass loss outside of $r_{1/2}$ reduces the cluster mass by ${\sim 30\%}$. Over this period, the cluster completes approximately 13 regular orbits around its host galaxy, with stable apocenters of $\approx 1.1~{\rm kpc}$ and pericenters of $\approx 200~{\rm pc}$. Subsequently, the mass loss accelerates and the cluster eventually dissolves after ${\sim} 2.5\,{\rm Gyr}$. The half-mass radius of the cluster remains nearly constant over the cluster's entire evolution. Although our simulations do not have the ability to accurately track the orbits of stellar particles over a Hubble time --- which would require a direct $N$-body integrator, not the softened force algorithm employed by {\sc Gizmo} --- and the force softening adopted here precludes the formation of clusters with ${\sim}{\rm pc}$-scale half-mass radii, these results indicate that the clusters forming in these simulations are long-lived and are plausible progenitors of present-day GCs that are observed in some dwarf galaxies. 

\subsection{Connections between stellar clusters and their host halos}
\label{subsec:galaxy_props}
The last 6 columns of Table~\ref{tab:data} contain information about the dark matter halo within which each cluster forms and the properties of the central galaxy of each halo at the time of cluster formation. As expected for halos that have $M_{\rm vir}(z=0) =10^{10}\,\msun$, the progenitor halos at high redshifts -- when the clusters form -- are approximately an order of magnitude lower in mass, $M{\sim} (0.5-1.5)\times 10^{9}\,\msun$, and the maximum circular velocity of the host at the time of cluster formation is $30-40\,{\rm km/s}$. These numbers indicate that these GCCs are forming at high redshift in halos that are a factor of ${\sim}10$ more massive than the atomic cooling limit of $T_{\rm vir}=10^4\,{\rm K} \leftrightarrow V_{\rm halo} = 17\,{\rm km/s}$ (or $M_{\rm vir} \approx 10^8\,\msun$~at $z {\sim} 8$). 

At the time the clusters form, the halos typically have very low stellar content: the \textit{most} massive ``galaxy" --- defined as the stellar mass within $0.1\,r_{\rm vir}$ prior to the formation of the cluster --- is $10^5\,\msun$, and in some cases, only $10^{4}\,\msun$ of stars exist near the center of the halo prior to the time of cluster formation. In this sense, the formation of the clusters is indeed pre-galactic, as there is essentially no galaxy present prior to the cluster formation episode. The pre-existing stellar populations tend to be fairly extended ($r_{1/2,\star} \approx 150-400\,{\rm pc}$) and metal-poor ($[{\rm Fe/H}] \sim -3)$.

\begin{figure*}
    \centering
    \includegraphics[width=0.95\textwidth]{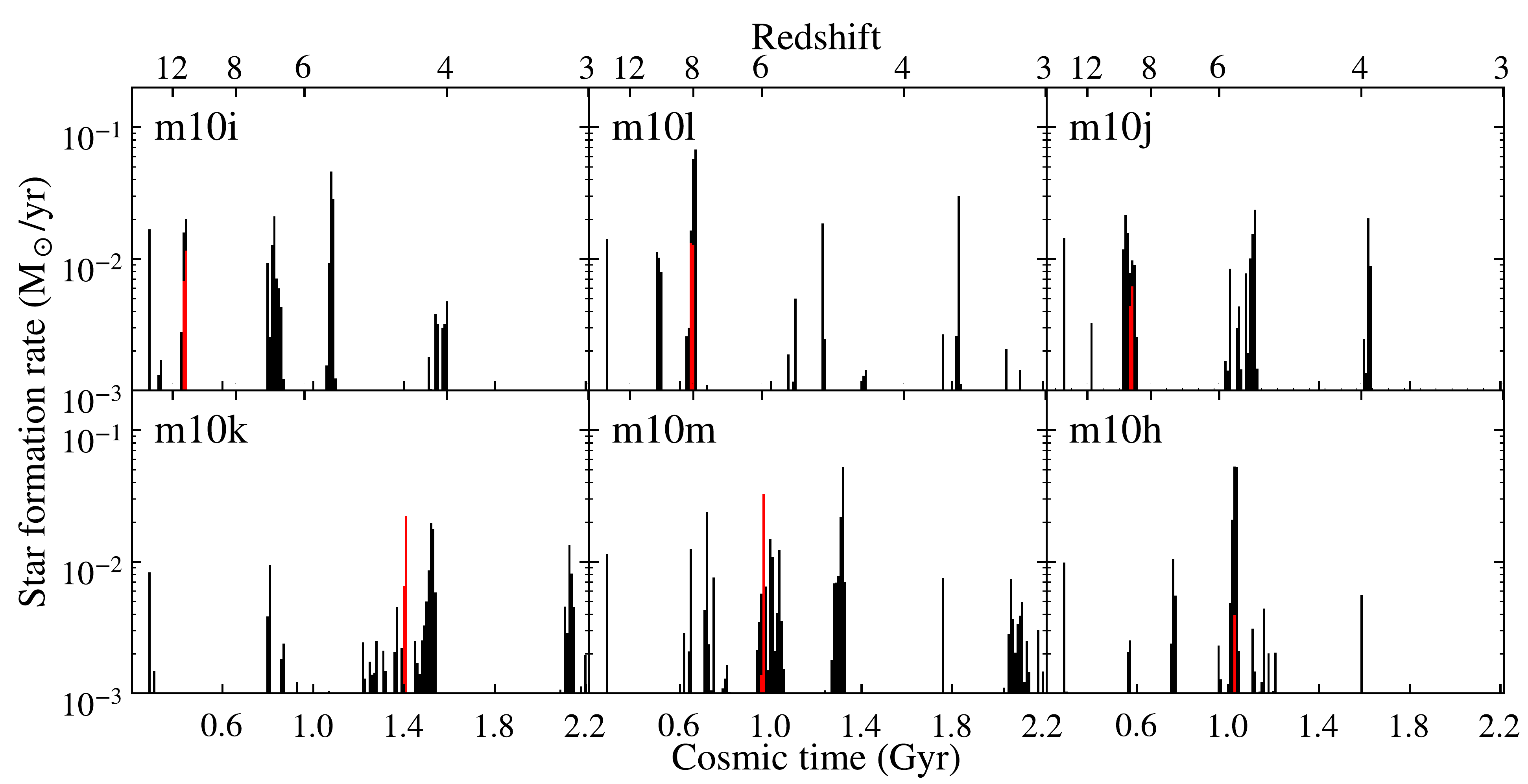}
     \includegraphics[width=0.95\textwidth]{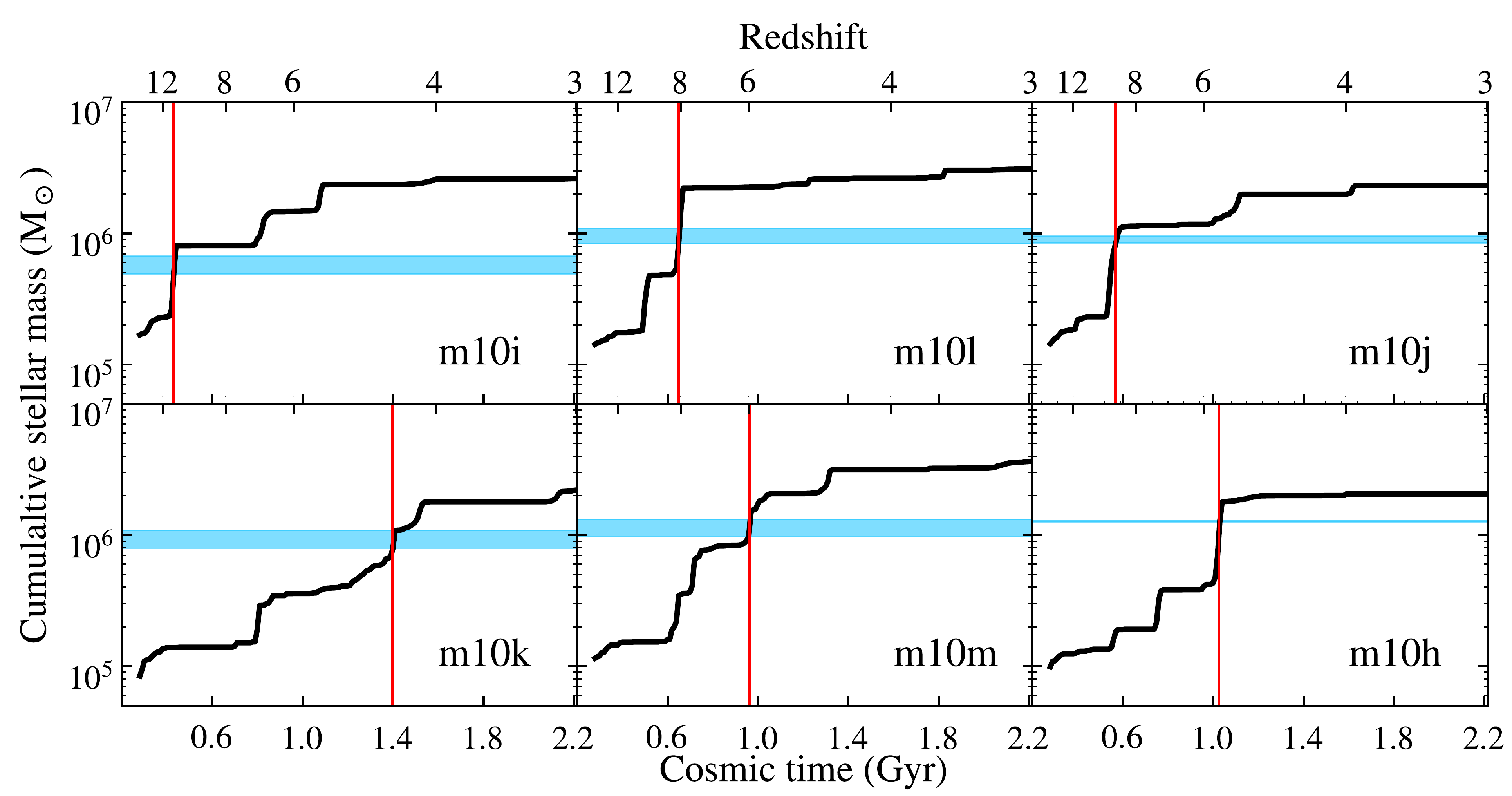}
    \caption{{\bf Top}: star formation rate within the virial radius of the main halo in each m10 realization with our fiducial FIRE2 physics, computed for 10 Myr time intervals. The formation of the stars formed in the massive bound clusters within each primary galaxy is plotted in red, while the black bars show all other star formation in the main halo. 
     {\bf Bottom}: the stellar mass assembly of the simulated galaxies over the cosmological evolution. The red vertical bands specify the cluster formation epochs for each realization, and the blue horizontal bands show the contribution of the cluster's mass to the archaeological star formation history of each primary galaxy.
     The star formation in these halos is very bursty and episodic, and the clusters often form at times of (relatively) intense but short-lived star formation. The star formation event leading to the formation of the cluster is often comprises most of the stellar content in the halo at the formation epoch, and in some cases, the cluster itself contains the majority of the stars within the halo at the time of its formation. We do not include the cluster that forms in the m10i box at $z=4.9$, as it forms outside of the virial radius of the most massive galaxy. 
     }
    \label{fig:sfr}
\end{figure*}

Fig. \ref{fig:sfr} shows the star formation rates (SFRs; top) and the archaeological star formation histories\footnote{Archaeological SFHs are computed by considering the formation times of all of the stars in the galaxy at $z=3$ as opposed to summing the instantaneous star formation rate in any individual progenitor.} (SFHs; bottom) of the halos in our fiducial suite; in both cases, we consider all stars within $r_{\rm vir}$ and use time bins of 10~Myr. 
We separate the SFR in each cluster (shown in red) from the rest of the stars in the halo (shown in black). In the SFH plots, we include both stars in clusters and all other stars, but we mark the cluster formation epoch (red vertical bands) and the clusters' contribution to the total SFHs (blue horizontal bands). The star formation in these simulated halos is highly episodic, as has been noted before for simulations both at this mass scale  \citep{stinson2007,dominguez2015,sparre2017,fitts2017,emami2019}: many have prolonged periods of true quiescence punctuated by brief periods of relatively vigorous star formation ($\dot{M}_{\star} \approx 0.01-0.1\,\msun\,{\rm yr}^{-1}$ for tens of Myr; see also Fig.~1 of \citealt{boylan-kolchin2015}). 

The clusters contribute substantially to the total stellar mass within the virial radius at the time of cluster formation (see also Table~\ref{tab:data}). In fact, for all but the least massive cluster we consider here (m10h), the clusters outweigh the pre-existing stellar mass within the virial radius at the time of cluster formation; the earliest-forming cluster (in m10i) is ten times more massive than the pre-existing stars in that halo. Even in m10h, the cluster has a mass that is equal to $50\%$ of the total stellar mass that was present in the halo prior to cluster formation. These results demonstrate that high-redshift star clusters in progenitors of present-day dwarf galaxies may contain a substantial --- or even dominant --- fraction of the total stellar mass in the halo at the time the clusters form. As seen in Figure~\ref{fig:sfr}, the clusters often form at times of star formation bursts in the dwarfs, which is not surprising given the importance of SN feedback in producing conditions conducive to star cluster formation (as discussed in Sec.~\ref{sec:formation-mechanism}). As a result, stars formed within ${\sim 50}$~Myr of the cluster formation epoch often account for over 10\% (and, in the case of m10l, almost 70\%) of all stars residing within the halo's virial radius at $z=3$. The formation of both of the clusters themselves and of stars associate with the clusters' birth clouds are defining events for the star formation histories of these simulated dwarf galaxies. We return to observational implications of high fraction of a halo's total stellar mass at the epoch of cluster formation attributable to the cluster in Sec.~\ref{subsec:observe}.

The formation sites of the GCs are also of substantial interest. Many ``pre-galactic" models for GCs -- or standard interpretations of these models -- assume that GCs form at the centers of dark matter mini-halos, leading to the prediction that GCs should be immersed in a dense dark matter cocoon that should be detectable in the kinematics at the outskirts of GCs \citep{peebles1984,heggie1996, mashchenko2005, conroy2011, ibata2013, penarrubia2017,boldrini2020}. However, dense molecular clouds do not necessarily form directly at the center of dark matter halos, and so it is not at all obvious that even pre-galactic GC models necessarily produce GCs that form directly at the minimum of the dark matter gravitational potential and bear the indelible imprints of dark matter hosts. It is natural to assume that GCs form near the centers of their host galaxies, as the galaxies by definition trace the bulk of the star formation in the halo, but we have shown the GCCs simulated here can contain comparable stellar content to the galaxy in the early epochs corresponding to the GC formation times for these dwarf galaxies (which also form in atomic cooling halos, not mini-halos).

In fact, as Table~\ref{tab:data} indicates, many of the simulated clusters form at substantial distances ($3-5\,r_{1/2,\rm gal}$ or more) from their ``host galaxies'' as defined by the pre-existing stars. While often there are connections between the galaxies and clusters, as evidenced by periods of increased star formation activity in the galaxy at the same time as cluster formation even for clusters that form well outside their host's stellar half-mass radius, the clusters are generally not deeply embedded within the nascent galaxies, even though the halos studied here all have cusped density profiles at their centers. This is likely due to a combination of the high redshift of formation, the (relatively) low masses of dwarf galaxy dark matter halos, and conditions necessary for cluster formation: dense gaseous regions in the turbulent ISM of dwarf galaxy progenitors can span several hundred pc (see Figures~\ref{fig:m10l} and \ref{fig:m10i}), significantly exceeding the typical size of any pre-existing ``galaxy". As a result, stochastic events that trigger cluster formation --- cloud collisions or compressive supernova shocks --- often occur at locations that do not coincide with the bulk of the existing stars. 

The large sizes (relative to the entire dark matter halo) of the GMC complexes that result in star clusters also results in clusters that do not typically form within the regions of the highest dark matter density, offering a natural explanation of the lack of observed dark matter around GCs even in scenarios where they form in low-mass halos at high redshifts. In all cases studied here, the clusters' mass is heavily dominated by stars, with dark matter contributing at most $15\%$ of the mass within $r_{1/2}$. This dark matter contribution --- equal to at most 10 dark matter particles --- is consistent with the expected dark matter contribution of the main halo at the clusters' formation location as opposed to these clusters forming at the centers of their own dark matter halos.

\section{Discussion}
\label{sec:discuss}
\subsection{Observational consequences}
\label{subsec:observe}
Given the recent advances in both theoretical and observational understanding of in situ star cluster formation in the distant Universe, along with the promise of forthcoming JWST observations, the potential observational implications of the results presented in Section~\ref{sec:clusters} are worth exploring. Most of the clusters in our current sample --- 4 of the primary 6 clusters --- attain star formation rates of $0.02-0.03~\msun\,{\rm yr}^{-1}$ (corresponding to $(2-3)\times 10^{5}\,\msun$ of stars formed in a 10~Myr window) at $z \sim 5-11$, which would result in an absolute luminosity of $M_{\rm UV}\approx -14$ mag ($m \approx 33$ at $z\approx 8$). Given the sizes of these simulated systems, $r_{1/2} \approx 20\,{\rm pc}$, the corresponding surface brightness would be roughly $\mu=22-23\,{\rm mag\,arcsec^2}$, with a star formation rate surface density of $\Sigma_{\rm SFR} \approx 10^{2}\,\msun\,{\rm yr^{-1}\,kpc^{-2}}$ for $\approx 10\,{\rm Myr}$. These properties are in line with both predictions from GC formation models \citep{ricotti2016, renzini2017, mbk2018} and observations of possible star clusters in formation at high redshift \citep{vanzella2017,vanzella2019,vanzella2021b}. 

The sizes of the clusters at formation --- $r_{1/2} \approx 20~{\rm pc}$ --- are consistent with the most compact sources with comparable luminosities ($M_{\rm UV} \approx -15$) seen in lensed \textit{Hubble Space Telescope} images \citep{vanzella2019,kikuchihara2020,bouwens2021a}. As Figures~\ref{fig:m10l}, \ref{fig:m10i}, and \ref{fig:sfr} demonstrate, there is typically substantial additional star formation that is coeval with cluster formation; accounting for this extra-cluster star formation, the total star formation rates can be $\approx 0.1\,\msun\,{\rm yr}^{-1}$ over an area of several hundred pc. This would result in an absolute luminosity of $M_{\rm UV}\approx -15.5$ mag ($m \approx 31.5$ at $z\approx 8$) for the full host system, which is at the edge of detectability in deep JWST fields. In lensing fields, the GCCs would likely stand out from the rest of the coeval star formation because of their much higher surface densities \citep{zick2018}.

Although all of the star clusters discussed here live for tens to hundreds of dynamical times, none survive to the present day. This is likely a reflection of the artificially large size of the clusters (owing to our adopted force softening) and the use of a TreePM code, rather than a code that directly implements gravitational interactions, for our simulations. In the future, we plan to use hybrid schemes to more faithfully track the evolution of clusters formed at high redshift in progenitors of present-day dwarf galaxies to see if the clusters are indeed analogues of the old, metal-poor clusters observed in some low-mass dwarfs. 

The results presented here are promising, if non-definitive, in this regard. Many properties of the clusters studied here are consistent with those expected of ancient globular clusters in Local Group dwarfs at earlier epochs, closer to their formation. In particular, the clusters here (1) contain a substantial fraction of the stars relative to the host galaxy; (2) have similar chemistry to the oldest stars in their host galaxies; and (3) form at locations that would likely allow them to survive to the present day. The conditions that lead to the formation of bound stellar clusters at high redshifts in our simulated dwarf galaxies therefore represent plausible formation scenarios for the growing population of observed ancient GCs in nearby dwarf galaxies. These clusters also form with the efficiency required for reproducing the observed relationship between globular cluster system mass ($M_{\rm GCs})$ and dark matter halo mass ($M_{\rm halo}$) in the local Universe of $M_{\rm GCs}/M_{\rm halo} \approx 4 \times 10^{-5}$ \citep{blakeslee1997, spitler2009, georgiev2010, hudson2014, harris2015, burkert2020}: as shown in \citet{mbk2017}, this is naturally achieved if a $M_{\star}=2\times 10^5\,\msun$ GC forms in a $M_{\rm halo}\approx 10^{9}\,\msun$ dark matter halo at $z \sim 8$, which is remarkably similar to what we find in this work. 

\subsection{Sensitivity to simulation choices}
\label{subsec:sensitivity}
The GCCs described in this paper form in very high surface density gas, often following compression caused by strong stellar feedback. It is therefore important to examine the effects of choices related to stellar feedback and star formation criteria in order to understand how robust our results are to details of the prescriptions we adopt. In this section, we explore the roles that work related to stellar mass-loss and the star formation density threshold play in driving our results.

As described and tested in detail in \citet{hopkins2018b}, we account for conversion of thermal energy into kinetic energy following supernova explosions (i.e., any ``$P\,dV$ work'' done) during the Sedov-Taylor phase of the expansion before the remnant reaches the resolution scale $\Delta x$ at which the coupling to the nearest gas cells occurs, which is crucial for producing converged results. However, as discussed there and in \citet{hopkins2018}, how to treat the ratio of thermal to kinetic energy injected for stellar mass-loss when the mass-loss is discretized into finite time-steps is more ambiguous if there is a continuously-expanding bubble below the resolution scale. Moreover, it is clear that the Sedov-Taylor solution, which assumes a single instantaneous discrete energy injection event, is not the correct solution for a continuous wind. In our fiducial simulations, therefore, we ignore any unresolved $P\,dV$ work from stellar mass-loss processes\footnote{We note that our treatment of stellar mass-loss processes is different than the standard FIRE-2 physics \citep{hopkins2018}, where 100\% of stellar mass-loss energy is converted into macroscopic momentum, but is consistent with the approach adopted in FIRE-3 \citep{hopkins2022}.}, which is consistent with both observations and recent simulations  \citep{harper2009,lancaster2021a,lancaster2021b}. 

However, it is interesting to ask what might happen if stellar mass-loss bubbles did undergo a prolonged energy-conserving phase during which substantial $P\,dV$ work was done, converting almost all of the thermalized/shocked ejecta energy into kinetic energy (momentum) on large scales. We do this by treating each stellar mass-loss event (which injects some $\Delta M \equiv \dot{M}_{\ast}\,\Delta t$, with initial free-streaming kinetic luminosity/energy $\Delta E \equiv \dot{E}\,\Delta t$) as a ``mini-supernova'' and applying the exact same treatment as we do for SNe following \citet{hopkins2018b}. The practical effect of this (given the various scaling for e.g. the cooling radii of SNe) is that most of total energy injection by stellar mass-loss is converted into momentum/kinetic energy on resolved scales (i.e.\ $\sim 100\%$ of the stellar mass-loss energy is converted into macroscopic momentum), as compared to post-shock thermal energy that can be more efficiently radiated away. We emphasize that the particular functional form we adopt has no clear physical motivation, but it provides a useful comparison point for understand possible effects of unresolved stellar mass loss physics on our results.

Surprisingly, the ``stronger'' stellar mass-loss given by this assumption produces effectively weaker SN feedback (higher SFRs and stellar masses). This is related to the effects shown and discussed in \citet{hopkins2020}. If ``early'' stellar feedback (processes that act before SNe explode in a young star-forming region, e.g. radiative heating and radiation pressure and stellar mass-loss) is much weaker, then that region collapses much further and produces many more stars in a denser configuration. When, approximately $40$ Myr later, those stars begin to explode, the SNe (which carry energy that is an order-of-magnitude larger than the energy attributable to stellar mass-loss) are much more strongly clustered, making it easier for bubbles to overlap and driving much stronger outflows (as has also been seen in idealized experiments that vary the strength of SNe clustering, e.g., \citealt{martizzi2015,walch2015,fielding2018}). When early feedback is artificially made much stronger as in our experiment, clouds are disrupted earlier with much lower star formation efficiencies, producing much weaker SNe clustering and therefore weaker {\em net} large-scale SN feedback. 

A comparison of SFHs between the ``early'' stellar feedback model and our fiducial suite shows that the fiducial suite simulations have formed slightly fewer stars, in agreement with the argument outlined above. Furthermore, {\it no} bound stellar clusters above our chosen threshold cluster mass of  $M_\star=5\times10^4~\msun$ form in the simulations with modified stellar mass-loss processes (stronger early feedback; less clustered SNe feedback). However, lower-mass clusters $(M_{\star} < 5\times 10^4\,\msun$) do form in the simulations with the alternate mass loss treatment (which is the default treatment in FIRE-2 but not in FIRE-3). These low-mass clusters are always destroyed quickly (within 20~Myr). \citet{kim2018} and \citet{Ma2020} both found that massive and long-lived clusters \textit{do} form using this alternate mass loss treatment in their simulations, which focus on significantly more massive systems ($M_{\rm halo} \gtrsim 10^{10}\,\msun$ at $z\sim 6$) that have star formation rates that are at least an order of magnitude higher than what we find here. This indicates that the larger-scale impact of the treatment of $P\,dV$ work from stellar mass loss depends on the mass of the cluster: higher-mass GMCs, which result in higher star formation rates, are less dependent on the treatment of unresolved $P\,dV$ work, likely because of their higher binding energies. One caveat to this conclusion is the relatively large force softening adopted in these simulations, which effectively limits the range of densities that star-forming gas can reach. 

A related point is that cluster formation efficiency is also sensitive to the star formation criteria adopted in simulations \citep{grudic2018,grudic2018b,Ma2020}. In our fiducial suite, we impose a star formation density threshold of $n_{\rm H}\geq n_{\rm thresh}=1000\, {\rm cm}^{-3}$ \citep{hopkins2018}. We have also performed a simulation of m10m where we change only the star formation criterion (to $n_{\rm thresh}=100\, {\rm cm}^{-3}$), and leave everything else unmodified from our fiducial simulation setup. In this simulation, stellar clusters form in a nearly identical manner as in the fiducial run. If we use this reduced density threshold, or alternately a flow convergence criteria, $\nabla \cdot {\rm v}<0$, for the velocity field of the gas cells \citep{grudic2018} instead of a density threshold criterion, along with the treatment of stellar mass-loss that is modified from our fiducial simulations, we find no bound clusters.  These results strongly indicate that the treatment of unresolved $P\,dV$ work from stellar mass-loss or other forms of unresolved early feedback is crucial in resolving the formation of star clusters in progenitors of $z=0$ dwarf galaxies while the choice of density threshold (if any) is not. Preliminary work using FIRE-3 galaxy formation physics \citep{hopkins2022}, which follows the treatment of unresolved $P\,dV$ work adopted here and does not adopt a density threshold for star formation --- it requires star-forming gas to be self-gravitating, Jeans unstable, and within a converging flow --- supports this conclusion (Sameie et al.~2022, in preparation).

\subsection{Comparison with previous results}
\label{subsec:previous}
While our work covers a different mass regime from what has previously been studied in full cosmological simulations that resolve the formation of GCCs (e.g., \citealt{kimm2016, kim2018, Ma2020}), our results are broadly consistent with those from these previous numerical simulation and analytic arguments: globular cluster candidates form preferentially at early cosmic times when the turbulent ISM of gas-rich galaxies can provide the requisite high pressures that are conducive to cluster formation.

Our results differ somewhat from those of semi-analytic models that aim to understand GC formation within the broader context of galaxy evolution across cosmic time. These models typically predict that GCs in $z=0$ dwarf galaxies form at relatively late times ($z \lesssim 3$), in part because the average metallicities of such galaxies are not predicted to reach the levels seen in most globular clusters until that point \citep{choksi2018,el-badry2019, reina-campos2019,kruijssen2019}. The clusters formed in our simulations form significantly earlier than this, at $z \gtrsim 5$. It is noteworthy that the GCCs found here have metallicities that are always at least as high as that of their host galaxies; in some cases, the GCCs are one full dex higher in metallicity. Allowing for GCs to have higher metallicities than the mean gas-phase metallicity of their progenitors could be an interesting and fruitful path forward for semi-analytic models of GC formation. 

Although we are unable to follow the evolution of the GCCs to $z=0$ and therefore cannot make definitive statements about whether these objects are truly GC analogs, many of their properties are grossly consistent with ancient GCs that are observed in some low-mass dwarf galaxies today. One clear difference from observations is the metallicity of the clusters in our simulations, which are slightly lower than metallicities of well-quantified GCs observed in dwarf galaxies \citep{beasley2019} and have larger iron metallicity spreads ($0.15-0.2$~dex; the metallicity distributions are reasonably well approximated by gaussians with $\sigma=0.15$~dex across the simulation suite) than are observed ($0.045$~dex; \citealt{carretta2009, bailin2019}). This could be an indication that the clusters formed here would be disrupted and form the streams that are known to have lower metallicities \citep{martin2022}, or it could mean that subtle aspects of the galaxy formation modeling in FIRE require refinements in this regime. Indeed, \citet{wheeler2019} noted that low-mass dwarf galaxies from FIRE-2 lie slightly but systematically below the observed mass-metallicity relationship. An explicit treatment of Population III star formation and enrichment is likely to be important for understanding the properties of the lowest metallicity systems and their connections to globular clusters (see also \citealt{schauer2021}).  

More broadly, the existence of galaxies such as Eridanus II, which formed 80\% of its stars before $z=6$ and hosts a GC with a stellar population indistinguishable from that of the galaxy \citep{simon2021}, demonstrates that GC formation in very low-mass systems at high redshifts ($z > 6$) is possible and must be accounted for in models. The properties of the GGs and galaxies in our simulated dwarf galaxy halos are in many ways similar to Eridanus II, with GC formation accompanying (or even preceding) the formation of the bulk of the stars in the galaxy, though our systems are somewhat higher in stellar mass. More detailed observational studies of GCs in dwarf galaxies, coupled with future simulations of such systems, hold the promise to reveal important aspects of star formation in metal-poor systems in the reionization era.

\section{Conclusions}
\label{sec:conclusion}
The recent discoveries of ancient GCs in low-mass ($M_{\star}\sim 10^5-10^7\,\msun$) Local Group dwarf galaxies and of star cluster candidates in formation in the high-redshift Universe has reignited interest in a number of questions related to GCs in dwarf galaxies. We have used cosmological hydrodynamic simulations of seven halos with virial masses of $M_{\rm vir}(z=0)\sim 10^{10}\, \text{M}_\odot$ from the FIRE-2 project to investigate star cluster formation in the ancestors of present-day dwarf galaxies. We find that star cluster formation at high redshift ($11 \gtrsim z \gtrsim 5$) is indeed common in these systems, which is perhaps the most important high-level result from our study. In more detail, our principal conclusions include the following:
\begin{itemize}
    \item Relatively massive ($M_{\star}\in [0.5-5]\times 10^5\,\msun$) and compact ($6 \lesssim r_{1/2} \lesssim 30\,{\rm pc}$) clusters form in halos with virial masses of $(0.5-2)\times 10^9\,\msun$ --- roughly a factor of ten more massive than the atomic cooling threshold corresponding to $T_{\rm vir}=10^4\,{\rm K}$ --- in the redshift range $11 \gtrsim z \gtrsim 5$. Of the seven systems studied here, five form one such cluster, while one forms two clusters and one forms no clusters.
    \item The clusters form when ${\sim}10^7\,\msun$ of dense and turbulent gas reaches a surface density of $\Sigma_{\rm thresh} = 10^4\,\msun\,{\rm pc}^{-2}$. These conditions occur because of compressive shocks from nearby star formation, cloud-cloud collisions, or both.
    \item Once the requisite conditions are achieved, cluster formation happens rapidly. Stellar feedback clears all gas from the cluster region by 10~Myr from the start of cluster star formation. In some cases, this feedback leads to shocks that trigger nearly coeval star formation hundreds of parsecs away from the cluster.
    \item The clusters and star formation associated with the GMCs from which the clusters form exceed the pre-existing stellar content of the halo, meaning the clusters form before there is a well-defined host galaxy at these mass scales. The GCCs studied here therefore originate from a phase of galaxy formation that either predates or accompanies the star formation constituting the bulk of the host galaxy.
    \item In several cases, the clusters constitute ${\sim} 10\%$ of the total stellar mass in the host halo at $z=3$. This is consistent with observations of GC-hosting dwarfs in the local Universe, where cluster stars typically constitute 1-10\% of the stellar mass of the galaxy \citep{georgiev2010, hudson2014, larsen2017}.
    \item The clusters in our simulations form at higher redshift ($11 \gtrsim z \gtrsim 5$) than is predicted at this mass scale in many semi-analytic models of cluster formation ($z \lesssim 3$; e.g., \citealt{choksi2018, el-badry2019, kruijssen2019}), in part because the models typically tie the mean gas-phase metallicity of a forming cluster to that of its host galaxy. By contrast, clusters in the simulations presented here form out of gas with higher-than-average metallicity (averaging over the gas in the halo at the cluster formation epoch) and often before a well-defined galaxy is even present. 
    \item The clusters live for tens to hundreds of dynamical times ($0.2-2.5~{\rm Gyr}$), with clusters born far from the dynamical center of the halo surviving the longest. The disruption of the clusters comes from a combination of physical and numerical effects; were it possible to accurately resolve the internal dynamics of the clusters, several might well survive to $z=0$.
    \item The clusters typically form outside of the half-light radius of any pre-existing galaxy (insofar as any such galaxy exists), well removed from the center of the host halo. In some cases, the cluster formation sites are at $0.25-0.5\,r_{\rm vir}$. These offsets are natural since clusters are forming in regions of dense gas having turbulent velocity dispersion comparable to the halo virial velocity, which makes the cluster formation sites somewhat stochastic.
    \item Given the formation sites, the clusters formed here are never deeply enshrouded in the centers of their own dark matter halos, though they all form within the virial radius of a $M_{\rm vir}\sim 10^9\,\msun$ dark matter halo. The contribution of dark matter to the clusters' mass profiles is minimal and consistent with the background halo density at the cluster formation location (typically hundreds of pc, or ${\sim}0.2\,r_{\rm vir}$, from the halo center).
    \item Properties of these clusters are consistent with objects detected in HST observations at $z \sim 6-8$ in lensing fields. They are also suggestive of similarities to clusters observed in Local Group dwarfs at $z=0$.
    \item The treatment of unresolved thermal energy deposition from stellar mass-loss is a primary numerical ambiguity affecting cluster formation physics in our suite.
    
\end{itemize}
While results from these ``proof-of-principle" simulations are both encouraging and intriguing, there are multiple avenues for improvement in the near term. Our choice of softening scale for baryonic particles, $2-5\,{\rm pc}$, leads to artificially large sizes for our clusters; running versions of these simulations with softenings that are roughly ten times smaller should allow us to test whether the sizes of the clusters we form are realistic. A more challenging issue is faithfully tracing the internal dynamics of the clusters for cosmological times (${\sim} 10$ or more Gyr). Hybrid numerical schemes that resolve cluster formation in cosmological simulations and then track cluster evolution with methods capable of tracking collisional stellar dynamics within a larger-scale galactic environment (e.g., Rodriguez et al., in preparation) are promising in this regard. Finally, the galaxy formation models employed here, which are part of the FIRE-2 suite, are subject to further refinement. Details of the treatment of various aspects of star formation will likely be important for accurately capturing the formation of bound, long-lived star clusters; updates incorporated into FIRE-3 \citep{hopkins2022} are a starting point in this direction. In the near future, however, it will be possible to run the kinds of simulations presented here with numerical parameters that allow us to form pc-scale clusters and to follow the evolution of the clusters and their host galaxies from birth in the high-redshift Universe to present day. Such simulations will be crucial for using JWST data to constrain models of the formation and evolution of GCs \citep{renaud2018,forbes2018,adamo2020}.

\section*{Data Availability}
The data supporting the plots within this article are available on reasonable request to the corresponding author.

\section*{Acknowledgements}
MBK acknowledges support from NSF CAREER award AST-1752913, NSF grants AST-1910346 and AST-2108962, NASA grant NNX17AG29G, and HST-AR-15006, HST-AR-15809, HST-GO-15658, HST-GO-15901, HST-GO-15902, HST-AR-16159, and HST-GO-16226 from the Space Telescope Science Institute (STScI), which is operated by AURA, Inc., under NASA contract NAS5-26555. Support for PFH was provided by NSF Research Grants 1911233, 20009234, 2108318, NSF CAREER grant 1455342, and NASA grants 80NSSC18K0562 \& HST-AR-15800. AW received support from: NSF grants CAREER 2045928 and 2107772; NASA ATP grant 80NSSC20K0513; HST grants AR-15809 and GO-15902 from STScI; a Scialog Award from the Heising-Simons Foundation; and a Hellman Fellowship. JSB was supported by NSF grant AST-1910346. EQ was supported in part by a Simons Investigator grant from the Simons Foundation and NSF AST grant 2107872. JS was supported by an NSF Astronomy and Astrophysics Postdoctoral Fellowship under award AST-2102729. DRW acknowledges support from HST-GO-15476, HST-GO-15901, HST-GO-15902, HST-AR-16159, and HST-GO-16226 from STScI.

This work used the Extreme Science and Engineering Discovery Environment (XSEDE; \citealt{xsede}), via allocation AST140080, and the Frontera computing project at the Texas Advanced Computing Center (via allocations AST21010 and AST20016), which are supported by National Science Foundation awards ACI-1548562 and OAC-1818253, respectively. The analysis in this paper is carried out by python packages {\sc Numpy} \citep{numpy2020}, {\sc matplotlib} \citep{hunter2007}, {\sc scipy} \citep{scipy2020}, and {\sc h5py} \citep{collette2013}. 




\bibliographystyle{mnras}
\bibliography{bib} 

\begin{thebibliography}{}
\makeatletter
\relax
\def\mn@urlcharsother{\let\do\@makeother \do\$\do\&\do\#\do\^\do\_\do\%\do\~}
\def\mn@doi{\begingroup\mn@urlcharsother \@ifnextchar [ {\mn@doi@}
  {\mn@doi@[]}}
\def\mn@doi@[#1]#2{\def\@tempa{#1}\ifx\@tempa\@empty \href
  {http://dx.doi.org/#2} {doi:#2}\else \href {http://dx.doi.org/#2} {#1}\fi
  \endgroup}
\def\mn@eprint#1#2{\mn@eprint@#1:#2::\@nil}
\def\mn@eprint@arXiv#1{\href {http://arxiv.org/abs/#1} {{\tt arXiv:#1}}}
\def\mn@eprint@dblp#1{\href {http://dblp.uni-trier.de/rec/bibtex/#1.xml}
  {dblp:#1}}
\def\mn@eprint@#1:#2:#3:#4\@nil{\def\@tempa {#1}\def\@tempb {#2}\def\@tempc
  {#3}\ifx \@tempc \@empty \let \@tempc \@tempb \let \@tempb \@tempa \fi \ifx
  \@tempb \@empty \def\@tempb {arXiv}\fi \@ifundefined
  {mn@eprint@\@tempb}{\@tempb:\@tempc}{\expandafter \expandafter \csname
  mn@eprint@\@tempb\endcsname \expandafter{\@tempc}}}

\bibitem[\protect\citeauthoryear{{Adamo} et~al.,}{{Adamo}
  et~al.}{2020}]{adamo2020}
{Adamo} A.,  et~al., 2020, \mn@doi [\ssr] {10.1007/s11214-020-00690-x}, \href
  {https://ui.adsabs.harvard.edu/abs/2020SSRv..216...69A} {216, 69}

\bibitem[\protect\citeauthoryear{{Ashman} \& {Zepf}}{{Ashman} \&
  {Zepf}}{1992}]{ashman1992}
{Ashman} K.~M.,  {Zepf} S.~E.,  1992, \mn@doi [\apj] {10.1086/170850}, \href
  {https://ui.adsabs.harvard.edu/abs/1992ApJ...384...50A} {384, 50}

\bibitem[\protect\citeauthoryear{{Bailin}}{{Bailin}}{2019}]{bailin2019}
{Bailin} J.,  2019, \mn@doi [\apjs] {10.3847/1538-4365/ab4812}, \href
  {https://ui.adsabs.harvard.edu/abs/2019ApJS..245....5B} {245, 5}

\bibitem[\protect\citeauthoryear{{Bastian}, {Hollyhead}  \&
  {Cabrera-Ziri}}{{Bastian} et~al.}{2014}]{bastian2014}
{Bastian} N.,  {Hollyhead} K.,   {Cabrera-Ziri} I.,  2014, \mn@doi [\mnras]
  {10.1093/mnras/stu1775}, \href
  {https://ui.adsabs.harvard.edu/abs/2014MNRAS.445..378B} {445, 378}

\bibitem[\protect\citeauthoryear{{Beasley}, {Leaman}, {Gallart}, {Larsen},
  {Battaglia}, {Monelli}  \& {Pedreros}}{{Beasley} et~al.}{2019}]{beasley2019}
{Beasley} M.~A.,  {Leaman} R.,  {Gallart} C.,  {Larsen} S.~S.,  {Battaglia} G.,
   {Monelli} M.,   {Pedreros} M.~H.,  2019, \mn@doi [\mnras]
  {10.1093/mnras/stz1349}, \href
  {https://ui.adsabs.harvard.edu/abs/2019MNRAS.487.1986B} {487, 1986}

\bibitem[\protect\citeauthoryear{{Bechtol} et~al.,}{{Bechtol}
  et~al.}{2015}]{bechtol2015}
{Bechtol} K.,  et~al., 2015, \mn@doi [\apj] {10.1088/0004-637X/807/1/50}, \href
  {https://ui.adsabs.harvard.edu/abs/2015ApJ...807...50B} {807, 50}

\bibitem[\protect\citeauthoryear{{Behroozi}, {Wechsler}  \& {Wu}}{{Behroozi}
  et~al.}{2013}]{behroozi2013}
{Behroozi} P.~S.,  {Wechsler} R.~H.,   {Wu} H.-Y.,  2013, \mn@doi [\apj]
  {10.1088/0004-637X/762/2/109}, \href
  {https://ui.adsabs.harvard.edu/abs/2013ApJ...762..109B} {762, 109}

\bibitem[\protect\citeauthoryear{{Benson}, {Lacey}, {Baugh}, {Cole}  \&
  {Frenk}}{{Benson} et~al.}{2002}]{benson2002}
{Benson} A.~J.,  {Lacey} C.~G.,  {Baugh} C.~M.,  {Cole} S.,   {Frenk} C.~S.,
  2002, \mn@doi [\mnras] {10.1046/j.1365-8711.2002.05387.x}, \href
  {https://ui.adsabs.harvard.edu/abs/2002MNRAS.333..156B} {333, 156}

\bibitem[\protect\citeauthoryear{{Blakeslee}, {Tonry}  \&
  {Metzger}}{{Blakeslee} et~al.}{1997}]{blakeslee1997}
{Blakeslee} J.~P.,  {Tonry} J.~L.,   {Metzger} M.~R.,  1997, \mn@doi [\aj]
  {10.1086/118488}, \href
  {https://ui.adsabs.harvard.edu/abs/1997AJ....114..482B} {114, 482}

\bibitem[\protect\citeauthoryear{{Bolatto}, {Leroy}, {Rosolowsky}, {Walter}  \&
  {Blitz}}{{Bolatto} et~al.}{2008}]{bolatto2008}
{Bolatto} A.~D.,  {Leroy} A.~K.,  {Rosolowsky} E.,  {Walter} F.,   {Blitz} L.,
  2008, \mn@doi [\apj] {10.1086/591513}, \href
  {https://ui.adsabs.harvard.edu/abs/2008ApJ...686..948B} {686, 948}

\bibitem[\protect\citeauthoryear{{Boldrini}, {Mohayaee}  \& {Silk}}{{Boldrini}
  et~al.}{2020}]{boldrini2020}
{Boldrini} P.,  {Mohayaee} R.,   {Silk} J.,  2020, \mn@doi [\mnras]
  {10.1093/mnras/staa011}, \href
  {https://ui.adsabs.harvard.edu/abs/2020MNRAS.492.3169B} {492, 3169}

\bibitem[\protect\citeauthoryear{{Bouwens}, {Illingworth}, {van Dokkum},
  {Oesch}, {Stefanon}  \& {Ribeiro}}{{Bouwens} et~al.}{2021a}]{bouwens2021b}
{Bouwens} R.~J.,  {Illingworth} G.~D.,  {van Dokkum} P.~G.,  {Oesch} P.~A.,
  {Stefanon} M.,   {Ribeiro} B.,  2021a, arXiv e-prints, \href
  {https://ui.adsabs.harvard.edu/abs/2021arXiv211202948B} {p. arXiv:2112.02948}

\bibitem[\protect\citeauthoryear{{Bouwens}, {Illingworth}, {van Dokkum},
  {Ribeiro}, {Oesch}  \& {Stefanon}}{{Bouwens} et~al.}{2021b}]{bouwens2021a}
{Bouwens} R.~J.,  {Illingworth} G.~D.,  {van Dokkum} P.~G.,  {Ribeiro} B.,
  {Oesch} P.~A.,   {Stefanon} M.,  2021b, \mn@doi [\aj]
  {10.3847/1538-3881/abfda6}, \href
  {https://ui.adsabs.harvard.edu/abs/2021AJ....162..255B} {162, 255}

\bibitem[\protect\citeauthoryear{{Boylan-Kolchin}}{{Boylan-Kolchin}}{2017}]{mbk2017}
{Boylan-Kolchin} M.,  2017, \mn@doi [\mnras] {10.1093/mnras/stx2164}, \href
  {https://ui.adsabs.harvard.edu/abs/2017MNRAS.472.3120B} {472, 3120}

\bibitem[\protect\citeauthoryear{{Boylan-Kolchin}}{{Boylan-Kolchin}}{2018}]{mbk2018}
{Boylan-Kolchin} M.,  2018, \mn@doi [\mnras] {10.1093/mnras/sty1490}, \href
  {https://ui.adsabs.harvard.edu/abs/2018MNRAS.479..332B} {479, 332}

\bibitem[\protect\citeauthoryear{{Boylan-Kolchin}, {Weisz}, {Johnson},
  {Bullock}, {Conroy}  \& {Fitts}}{{Boylan-Kolchin}
  et~al.}{2015}]{boylan-kolchin2015}
{Boylan-Kolchin} M.,  {Weisz} D.~R.,  {Johnson} B.~D.,  {Bullock} J.~S.,
  {Conroy} C.,   {Fitts} A.,  2015, \mn@doi [\mnras] {10.1093/mnras/stv1736},
  \href {https://ui.adsabs.harvard.edu/abs/2015MNRAS.453.1503B} {453, 1503}

\bibitem[\protect\citeauthoryear{{Bullock}, {Kravtsov}  \&
  {Weinberg}}{{Bullock} et~al.}{2000}]{bullock2000}
{Bullock} J.~S.,  {Kravtsov} A.~V.,   {Weinberg} D.~H.,  2000, \mn@doi [\apj]
  {10.1086/309279}, \href
  {https://ui.adsabs.harvard.edu/abs/2000ApJ...539..517B} {539, 517}

\bibitem[\protect\citeauthoryear{{Burkert} \& {Forbes}}{{Burkert} \&
  {Forbes}}{2020}]{burkert2020}
{Burkert} A.,  {Forbes} D.~A.,  2020, \mn@doi [\aj] {10.3847/1538-3881/ab5b0e},
  \href {https://ui.adsabs.harvard.edu/abs/2020AJ....159...56B} {159, 56}

\bibitem[\protect\citeauthoryear{{Carlberg}}{{Carlberg}}{2002}]{carlberg2002}
{Carlberg} R.~G.,  2002, \mn@doi [\apj] {10.1086/340500}, \href
  {https://ui.adsabs.harvard.edu/abs/2002ApJ...573...60C} {573, 60}

\bibitem[\protect\citeauthoryear{{Carlberg}}{{Carlberg}}{2020}]{carlberg2020}
{Carlberg} R.~G.,  2020, \mn@doi [\apj] {10.3847/1538-4357/ab80bf}, \href
  {https://ui.adsabs.harvard.edu/abs/2020ApJ...893..116C} {893, 116}

\bibitem[\protect\citeauthoryear{{Carretta}, {Bragaglia}, {Gratton}, {D'Orazi}
  \& {Lucatello}}{{Carretta} et~al.}{2009}]{carretta2009}
{Carretta} E.,  {Bragaglia} A.,  {Gratton} R.,  {D'Orazi} V.,   {Lucatello} S.,
   2009, \mn@doi [\aap] {10.1051/0004-6361/200913003}, \href
  {https://ui.adsabs.harvard.edu/abs/2009A&A...508..695C} {508, 695}

\bibitem[\protect\citeauthoryear{{Choksi}, {Gnedin}  \& {Li}}{{Choksi}
  et~al.}{2018}]{choksi2018}
{Choksi} N.,  {Gnedin} O.~Y.,   {Li} H.,  2018, \mn@doi [\mnras]
  {10.1093/mnras/sty1952}, \href
  {https://ui.adsabs.harvard.edu/abs/2018MNRAS.480.2343C} {480, 2343}

\bibitem[\protect\citeauthoryear{{Col{\'\i}n}, {V{\'a}zquez-Semadeni}  \&
  {G{\'o}mez}}{{Col{\'\i}n} et~al.}{2013}]{colin2013}
{Col{\'\i}n} P.,  {V{\'a}zquez-Semadeni} E.,   {G{\'o}mez} G.~C.,  2013,
  \mn@doi [\mnras] {10.1093/mnras/stt1409}, \href
  {https://ui.adsabs.harvard.edu/abs/2013MNRAS.435.1701C} {435, 1701}

\bibitem[\protect\citeauthoryear{Collette}{Collette}{2013}]{collette2013}
Collette A.,  2013, Python and HDF5: unlocking scientific data.
" O'Reilly Media, Inc."

\bibitem[\protect\citeauthoryear{{Conroy}, {Loeb}  \& {Spergel}}{{Conroy}
  et~al.}{2011}]{conroy2011}
{Conroy} C.,  {Loeb} A.,   {Spergel} D.~N.,  2011, \mn@doi [\apj]
  {10.1088/0004-637X/741/2/72}, \href
  {https://ui.adsabs.harvard.edu/abs/2011ApJ...741...72C} {741, 72}

\bibitem[\protect\citeauthoryear{{Creasey}, {Sales}, {Peng}  \&
  {Sameie}}{{Creasey} et~al.}{2019}]{creasey2019}
{Creasey} P.,  {Sales} L.~V.,  {Peng} E.~W.,   {Sameie} O.,  2019, \mn@doi
  [\mnras] {10.1093/mnras/sty2701}, \href
  {https://ui.adsabs.harvard.edu/abs/2019MNRAS.482..219C} {482, 219}

\bibitem[\protect\citeauthoryear{{Crnojevi{\'c}}, {Sand}, {Zaritsky},
  {Spekkens}, {Willman}  \& {Hargis}}{{Crnojevi{\'c}}
  et~al.}{2016}]{crnojevic2016}
{Crnojevi{\'c}} D.,  {Sand} D.~J.,  {Zaritsky} D.,  {Spekkens} K.,  {Willman}
  B.,   {Hargis} J.~R.,  2016, \mn@doi [\apjl] {10.3847/2041-8205/824/1/L14},
  \href {https://ui.adsabs.harvard.edu/abs/2016ApJ...824L..14C} {824, L14}

\bibitem[\protect\citeauthoryear{{Dom{\'\i}nguez}, {Siana}, {Brooks},
  {Christensen}, {Bruzual}, {Stark}  \& {Alavi}}{{Dom{\'\i}nguez}
  et~al.}{2015}]{dominguez2015}
{Dom{\'\i}nguez} A.,  {Siana} B.,  {Brooks} A.~M.,  {Christensen} C.~R.,
  {Bruzual} G.,  {Stark} D.~P.,   {Alavi} A.,  2015, \mn@doi [\mnras]
  {10.1093/mnras/stv1001}, \href
  {https://ui.adsabs.harvard.edu/abs/2015MNRAS.451..839D} {451, 839}

\bibitem[\protect\citeauthoryear{{Eadie}, {Harris}  \& {Springford}}{{Eadie}
  et~al.}{2021}]{eadie2021}
{Eadie} G.~M.,  {Harris} W.~E.,   {Springford} A.,  2021, arXiv e-prints, \href
  {https://ui.adsabs.harvard.edu/abs/2021arXiv211015376E} {p. arXiv:2110.15376}

\bibitem[\protect\citeauthoryear{{El-Badry}, {Quataert}, {Weisz}, {Choksi}  \&
  {Boylan-Kolchin}}{{El-Badry} et~al.}{2019}]{el-badry2019}
{El-Badry} K.,  {Quataert} E.,  {Weisz} D.~R.,  {Choksi} N.,   {Boylan-Kolchin}
  M.,  2019, \mn@doi [\mnras] {10.1093/mnras/sty3007}, \href
  {https://ui.adsabs.harvard.edu/abs/2019MNRAS.482.4528E} {482, 4528}

\bibitem[\protect\citeauthoryear{{Elmegreen}}{{Elmegreen}}{2018}]{elmegreen2018}
{Elmegreen} B.~G.,  2018, \mn@doi [\apj] {10.3847/1538-4357/aaed45}, \href
  {https://ui.adsabs.harvard.edu/abs/2018ApJ...869..119E} {869, 119}

\bibitem[\protect\citeauthoryear{{Elmegreen} \& {Efremov}}{{Elmegreen} \&
  {Efremov}}{1997}]{elmegreen1997}
{Elmegreen} B.~G.,  {Efremov} Y.~N.,  1997, \mn@doi [\apj] {10.1086/303966},
  \href {https://ui.adsabs.harvard.edu/abs/1997ApJ...480..235E} {480, 235}

\bibitem[\protect\citeauthoryear{{Emami}, {Siana}, {Weisz}, {Johnson}, {Ma}  \&
  {El-Badry}}{{Emami} et~al.}{2019}]{emami2019}
{Emami} N.,  {Siana} B.,  {Weisz} D.~R.,  {Johnson} B.~D.,  {Ma} X.,
  {El-Badry} K.,  2019, \mn@doi [\apj] {10.3847/1538-4357/ab211a}, \href
  {https://ui.adsabs.harvard.edu/abs/2019ApJ...881...71E} {881, 71}

\bibitem[\protect\citeauthoryear{{Faucher-Gigu{\`e}re}, {Lidz}, {Zaldarriaga}
  \& {Hernquist}}{{Faucher-Gigu{\`e}re} et~al.}{2009}]{fg2009}
{Faucher-Gigu{\`e}re} C.-A.,  {Lidz} A.,  {Zaldarriaga} M.,   {Hernquist} L.,
  2009, \mn@doi [\apj] {10.1088/0004-637X/703/2/1416}, \href
  {https://ui.adsabs.harvard.edu/abs/2009ApJ...703.1416F} {703, 1416}

\bibitem[\protect\citeauthoryear{{Fielding}, {Quataert}  \&
  {Martizzi}}{{Fielding} et~al.}{2018}]{fielding2018}
{Fielding} D.,  {Quataert} E.,   {Martizzi} D.,  2018, \mn@doi [\mnras]
  {10.1093/mnras/sty2466}, \href
  {https://ui.adsabs.harvard.edu/abs/2018MNRAS.481.3325F} {481, 3325}

\bibitem[\protect\citeauthoryear{{Fitts} et~al.,}{{Fitts}
  et~al.}{2017}]{fitts2017}
{Fitts} A.,  et~al., 2017, \mn@doi [\mnras] {10.1093/mnras/stx1757}, \href
  {https://ui.adsabs.harvard.edu/abs/2017MNRAS.471.3547F} {471, 3547}

\bibitem[\protect\citeauthoryear{{Forbes} et~al.,}{{Forbes}
  et~al.}{2018}]{forbes2018}
{Forbes} D.~A.,  et~al., 2018, \mn@doi [Proceedings of the Royal Society of
  London Series A] {10.1098/rspa.2017.0616}, \href
  {https://ui.adsabs.harvard.edu/abs/2018RSPSA.47470616F} {474, 20170616}

\bibitem[\protect\citeauthoryear{{Geen}, {Soler}  \& {Hennebelle}}{{Geen}
  et~al.}{2017}]{geen2017}
{Geen} S.,  {Soler} J.~D.,   {Hennebelle} P.,  2017, \mn@doi [\mnras]
  {10.1093/mnras/stx1765}, \href
  {https://ui.adsabs.harvard.edu/abs/2017MNRAS.471.4844G} {471, 4844}

\bibitem[\protect\citeauthoryear{{Georgiev}, {Puzia}, {Goudfrooij}  \&
  {Hilker}}{{Georgiev} et~al.}{2010}]{georgiev2010}
{Georgiev} I.~Y.,  {Puzia} T.~H.,  {Goudfrooij} P.,   {Hilker} M.,  2010,
  \mn@doi [\mnras] {10.1111/j.1365-2966.2010.16802.x}, \href
  {https://ui.adsabs.harvard.edu/abs/2010MNRAS.406.1967G} {406, 1967}

\bibitem[\protect\citeauthoryear{{Grudi{\'c}}, {Hopkins},
  {Faucher-Gigu{\`e}re}, {Quataert}, {Murray}  \& {Kere{\v{s}}}}{{Grudi{\'c}}
  et~al.}{2018a}]{grudic2018}
{Grudi{\'c}} M.~Y.,  {Hopkins} P.~F.,  {Faucher-Gigu{\`e}re} C.-A.,  {Quataert}
  E.,  {Murray} N.,   {Kere{\v{s}}} D.,  2018a, \mn@doi [\mnras]
  {10.1093/mnras/sty035}, \href
  {https://ui.adsabs.harvard.edu/abs/2018MNRAS.475.3511G} {475, 3511}

\bibitem[\protect\citeauthoryear{{Grudi{\'c}}, {Guszejnov}, {Hopkins},
  {Lamberts}, {Boylan-Kolchin}, {Murray}  \& {Schmitz}}{{Grudi{\'c}}
  et~al.}{2018b}]{grudic2018b}
{Grudi{\'c}} M.~Y.,  {Guszejnov} D.,  {Hopkins} P.~F.,  {Lamberts} A.,
  {Boylan-Kolchin} M.,  {Murray} N.,   {Schmitz} D.,  2018b, \mn@doi [\mnras]
  {10.1093/mnras/sty2303}, \href
  {https://ui.adsabs.harvard.edu/abs/2018MNRAS.481..688G} {481, 688}

\bibitem[\protect\citeauthoryear{{Hahn} \& {Abel}}{{Hahn} \&
  {Abel}}{2011}]{hahn2011}
{Hahn} O.,  {Abel} T.,  2011, \mn@doi [\mnras]
  {10.1111/j.1365-2966.2011.18820.x}, \href
  {https://ui.adsabs.harvard.edu/abs/2011MNRAS.415.2101H} {415, 2101}

\bibitem[\protect\citeauthoryear{{Halbesma}, {Grand}, {G{\'o}mez}, {Marinacci},
  {Pakmor}, {Trick}, {Busch}  \& {White}}{{Halbesma}
  et~al.}{2020}]{halbesma2020}
{Halbesma} T. L.~R.,  {Grand} R. J.~J.,  {G{\'o}mez} F.~A.,  {Marinacci} F.,
  {Pakmor} R.,  {Trick} W.~H.,  {Busch} P.,   {White} S. D.~M.,  2020, \mn@doi
  [\mnras] {10.1093/mnras/staa1380}, \href
  {https://ui.adsabs.harvard.edu/abs/2020MNRAS.496..638H} {496, 638}

\bibitem[\protect\citeauthoryear{{Harper-Clark} \& {Murray}}{{Harper-Clark} \&
  {Murray}}{2009}]{harper2009}
{Harper-Clark} E.,  {Murray} N.,  2009, \mn@doi [\apj]
  {10.1088/0004-637X/693/2/1696}, \href
  {https://ui.adsabs.harvard.edu/abs/2009ApJ...693.1696H} {693, 1696}

\bibitem[\protect\citeauthoryear{{Harris}, {Harris}  \& {Hudson}}{{Harris}
  et~al.}{2015}]{harris2015}
{Harris} W.~E.,  {Harris} G.~L.,   {Hudson} M.~J.,  2015, \mn@doi [\apj]
  {10.1088/0004-637X/806/1/36}, \href
  {https://ui.adsabs.harvard.edu/abs/2015ApJ...806...36H} {806, 36}

\bibitem[\protect\citeauthoryear{{Harris} et~al.,}{{Harris}
  et~al.}{2020}]{numpy2020}
{Harris} C.~R.,  et~al., 2020, \mn@doi [\nat] {10.1038/s41586-020-2649-2},
  \href {https://ui.adsabs.harvard.edu/abs/2020Natur.585..357H} {585, 357}

\bibitem[\protect\citeauthoryear{{He}, {Ricotti}  \& {Geen}}{{He}
  et~al.}{2019}]{he2019}
{He} C.-C.,  {Ricotti} M.,   {Geen} S.,  2019, \mn@doi [\mnras]
  {10.1093/mnras/stz2239}, \href
  {https://ui.adsabs.harvard.edu/abs/2019MNRAS.489.1880H} {489, 1880}

\bibitem[\protect\citeauthoryear{{Heggie} \& {Hut}}{{Heggie} \&
  {Hut}}{1996}]{heggie1996}
{Heggie} D.~C.,  {Hut} P.,  1996, in {Hut} P.,  {Makino} J.,  eds,  IAU
  Symposium Vol. 174, Dynamical Evolution of Star Clusters: Confrontation of
  Theory and Observations. p.~303 (\mn@eprint {arXiv} {astro-ph/9511115})

\bibitem[\protect\citeauthoryear{{Herschel}}{{Herschel}}{1814}]{herschel1814}
{Herschel} W.,  1814, Philosophical Transactions of the Royal Society of London
  Series I, \href {https://ui.adsabs.harvard.edu/abs/1814RSPT..104..248H} {104,
  248}

\bibitem[\protect\citeauthoryear{{Hislop}, {Naab}, {Steinwandel}, {Lah{\'e}n},
  {Irodotou}, {Johansson}  \& {Walch}}{{Hislop} et~al.}{2022}]{hislop2022}
{Hislop} J.~M.,  {Naab} T.,  {Steinwandel} U.~P.,  {Lah{\'e}n} N.,  {Irodotou}
  D.,  {Johansson} P.~H.,   {Walch} S.,  2022, \mn@doi [\mnras]
  {10.1093/mnras/stab3347}, \href
  {https://ui.adsabs.harvard.edu/abs/2022MNRAS.509.5938H} {509, 5938}

\bibitem[\protect\citeauthoryear{{Hollyhead}, {Bastian}, {Adamo},
  {Silva-Villa}, {Dale}, {Ryon}  \& {Gazak}}{{Hollyhead}
  et~al.}{2015}]{hollyhead2015}
{Hollyhead} K.,  {Bastian} N.,  {Adamo} A.,  {Silva-Villa} E.,  {Dale} J.,
  {Ryon} J.~E.,   {Gazak} Z.,  2015, \mn@doi [\mnras] {10.1093/mnras/stv331},
  \href {https://ui.adsabs.harvard.edu/abs/2015MNRAS.449.1106H} {449, 1106}

\bibitem[\protect\citeauthoryear{{Hopkins}, {Kere{\v{s}}}, {O{\~n}orbe},
  {Faucher-Gigu{\`e}re}, {Quataert}, {Murray}  \& {Bullock}}{{Hopkins}
  et~al.}{2014}]{hopkins2014}
{Hopkins} P.~F.,  {Kere{\v{s}}} D.,  {O{\~n}orbe} J.,  {Faucher-Gigu{\`e}re}
  C.-A.,  {Quataert} E.,  {Murray} N.,   {Bullock} J.~S.,  2014, \mn@doi
  [\mnras] {10.1093/mnras/stu1738}, \href
  {https://ui.adsabs.harvard.edu/abs/2014MNRAS.445..581H} {445, 581}

\bibitem[\protect\citeauthoryear{{Hopkins} et~al.,}{{Hopkins}
  et~al.}{2018a}]{hopkins2018b}
{Hopkins} P.~F.,  et~al., 2018a, \mn@doi [\mnras] {10.1093/mnras/sty674}, \href
  {https://ui.adsabs.harvard.edu/abs/2018MNRAS.477.1578H} {477, 1578}

\bibitem[\protect\citeauthoryear{{Hopkins} et~al.,}{{Hopkins}
  et~al.}{2018b}]{hopkins2018}
{Hopkins} P.~F.,  et~al., 2018b, \mn@doi [\mnras] {10.1093/mnras/sty1690},
  \href {https://ui.adsabs.harvard.edu/abs/2018MNRAS.480..800H} {480, 800}

\bibitem[\protect\citeauthoryear{{Hopkins}, {Grudi{\'c}}, {Wetzel},
  {Kere{\v{s}}}, {Faucher-Gigu{\`e}re}, {Ma}, {Murray}  \& {Butcher}}{{Hopkins}
  et~al.}{2020}]{hopkins2020}
{Hopkins} P.~F.,  {Grudi{\'c}} M.~Y.,  {Wetzel} A.,  {Kere{\v{s}}} D.,
  {Faucher-Gigu{\`e}re} C.-A.,  {Ma} X.,  {Murray} N.,   {Butcher} N.,  2020,
  \mn@doi [\mnras] {10.1093/mnras/stz3129}, \href
  {https://ui.adsabs.harvard.edu/abs/2020MNRAS.491.3702H} {491, 3702}

\bibitem[\protect\citeauthoryear{{Hopkins} et~al.,}{{Hopkins}
  et~al.}{2022}]{hopkins2022}
{Hopkins} P.~F.,  et~al., 2022, arXiv e-prints, \href
  {https://ui.adsabs.harvard.edu/abs/2022arXiv220300040H} {p. arXiv:2203.00040}

\bibitem[\protect\citeauthoryear{{Hudson}, {Harris}  \& {Harris}}{{Hudson}
  et~al.}{2014}]{hudson2014}
{Hudson} M.~J.,  {Harris} G.~L.,   {Harris} W.~E.,  2014, \mn@doi [\apjl]
  {10.1088/2041-8205/787/1/L5}, \href
  {https://ui.adsabs.harvard.edu/abs/2014ApJ...787L...5H} {787, L5}

\bibitem[\protect\citeauthoryear{{Hunter}}{{Hunter}}{2007}]{hunter2007}
{Hunter} J.~D.,  2007, \mn@doi [Computing in Science and Engineering]
  {10.1109/MCSE.2007.55}, \href
  {https://ui.adsabs.harvard.edu/abs/2007CSE.....9...90H} {9, 90}

\bibitem[\protect\citeauthoryear{{Ibata}, {Nipoti}, {Sollima}, {Bellazzini},
  {Chapman}  \& {Dalessandro}}{{Ibata} et~al.}{2013}]{ibata2013}
{Ibata} R.,  {Nipoti} C.,  {Sollima} A.,  {Bellazzini} M.,  {Chapman} S.~C.,
  {Dalessandro} E.,  2013, \mn@doi [\mnras] {10.1093/mnras/sts302}, \href
  {https://ui.adsabs.harvard.edu/abs/2013MNRAS.428.3648I} {428, 3648}

\bibitem[\protect\citeauthoryear{{Ishigaki}, {Tominaga}, {Kobayashi}  \&
  {Nomoto}}{{Ishigaki} et~al.}{2018}]{ishigaki2018}
{Ishigaki} M.~N.,  {Tominaga} N.,  {Kobayashi} C.,   {Nomoto} K.,  2018,
  \mn@doi [\apj] {10.3847/1538-4357/aab3de}, \href
  {https://ui.adsabs.harvard.edu/abs/2018ApJ...857...46I} {857, 46}

\bibitem[\protect\citeauthoryear{{Johnson} et~al.,}{{Johnson}
  et~al.}{2017}]{johnson2017}
{Johnson} T.~L.,  et~al., 2017, \mn@doi [\apjl] {10.3847/2041-8213/aa7516},
  \href {https://ui.adsabs.harvard.edu/abs/2017ApJ...843L..21J} {843, L21}

\bibitem[\protect\citeauthoryear{{Katz} \& {Ricotti}}{{Katz} \&
  {Ricotti}}{2014}]{katz2014}
{Katz} H.,  {Ricotti} M.,  2014, \mn@doi [\mnras] {10.1093/mnras/stu1489},
  \href {https://ui.adsabs.harvard.edu/abs/2014MNRAS.444.2377K} {444, 2377}

\bibitem[\protect\citeauthoryear{{Katz} \& {White}}{{Katz} \&
  {White}}{1993}]{katz1993}
{Katz} N.,  {White} S. D.~M.,  1993, \mn@doi [\apj] {10.1086/172935}, \href
  {https://ui.adsabs.harvard.edu/abs/1993ApJ...412..455K} {412, 455}

\bibitem[\protect\citeauthoryear{{Kikuchihara} et~al.,}{{Kikuchihara}
  et~al.}{2020}]{kikuchihara2020}
{Kikuchihara} S.,  et~al., 2020, \mn@doi [\apj] {10.3847/1538-4357/ab7dbe},
  \href {https://ui.adsabs.harvard.edu/abs/2020ApJ...893...60K} {893, 60}

\bibitem[\protect\citeauthoryear{{Kim} et~al.,}{{Kim} et~al.}{2018}]{kim2018}
{Kim} J.-h.,  et~al., 2018, \mn@doi [\mnras] {10.1093/mnras/stx2994}, \href
  {https://ui.adsabs.harvard.edu/abs/2018MNRAS.474.4232K} {474, 4232}

\bibitem[\protect\citeauthoryear{{Kimm}, {Cen}, {Rosdahl}  \& {Yi}}{{Kimm}
  et~al.}{2016}]{kimm2016}
{Kimm} T.,  {Cen} R.,  {Rosdahl} J.,   {Yi} S.~K.,  2016, \mn@doi [\apj]
  {10.3847/0004-637X/823/1/52}, \href
  {https://ui.adsabs.harvard.edu/abs/2016ApJ...823...52K} {823, 52}

\bibitem[\protect\citeauthoryear{{Komatsu} et~al.,}{{Komatsu}
  et~al.}{2011}]{komatsu2011}
{Komatsu} E.,  et~al., 2011, \mn@doi [\apjs] {10.1088/0067-0049/192/2/18},
  \href {https://ui.adsabs.harvard.edu/abs/2011ApJS..192...18K} {192, 18}

\bibitem[\protect\citeauthoryear{{Koposov}, {Belokurov}, {Torrealba}  \&
  {Evans}}{{Koposov} et~al.}{2015}]{koposov2015}
{Koposov} S.~E.,  {Belokurov} V.,  {Torrealba} G.,   {Evans} N.~W.,  2015,
  \mn@doi [\apj] {10.1088/0004-637X/805/2/130}, \href
  {https://ui.adsabs.harvard.edu/abs/2015ApJ...805..130K} {805, 130}

\bibitem[\protect\citeauthoryear{{Kroupa}}{{Kroupa}}{2002}]{kroupa2002}
{Kroupa} P.,  2002, \mn@doi [Science] {10.1126/science.1067524}, \href
  {https://ui.adsabs.harvard.edu/abs/2002Sci...295...82K} {295, 82}

\bibitem[\protect\citeauthoryear{{Kruijssen}}{{Kruijssen}}{2019}]{kruijssen2019}
{Kruijssen} J.~M.~D.,  2019, \mn@doi [\mnras] {10.1093/mnrasl/slz052}, \href
  {https://ui.adsabs.harvard.edu/abs/2019MNRAS.486L..20K} {486, L20}

\bibitem[\protect\citeauthoryear{{Krumholz}, {McKee}  \&
  {Bland-Hawthorn}}{{Krumholz} et~al.}{2019}]{krumholz2019}
{Krumholz} M.~R.,  {McKee} C.~F.,   {Bland-Hawthorn} J.,  2019, \mn@doi [\araa]
  {10.1146/annurev-astro-091918-104430}, \href
  {https://ui.adsabs.harvard.edu/abs/2019ARA&A..57..227K} {57, 227}

\bibitem[\protect\citeauthoryear{{Lah{\'e}n}, {Naab}, {Johansson}, {Elmegreen},
  {Hu}, {Walch}, {Steinwandel}  \& {Moster}}{{Lah{\'e}n}
  et~al.}{2020}]{lahen2020}
{Lah{\'e}n} N.,  {Naab} T.,  {Johansson} P.~H.,  {Elmegreen} B.,  {Hu} C.-Y.,
  {Walch} S.,  {Steinwandel} U.~P.,   {Moster} B.~P.,  2020, \mn@doi [\apj]
  {10.3847/1538-4357/ab7190}, \href
  {https://ui.adsabs.harvard.edu/abs/2020ApJ...891....2L} {891, 2}

\bibitem[\protect\citeauthoryear{{Lah{\'e}n}, {Naab}  \&
  {Kauffmann}}{{Lah{\'e}n} et~al.}{2021}]{lahen2021}
{Lah{\'e}n} N.,  {Naab} T.,   {Kauffmann} G.,  2021, arXiv e-prints, \href
  {https://ui.adsabs.harvard.edu/abs/2021arXiv211114875L} {p. arXiv:2111.14875}

\bibitem[\protect\citeauthoryear{{Lake}, {Naoz}, {Chiou}, {Burkhart},
  {Marinacci}, {Vogelsberger}  \& {Kremer}}{{Lake} et~al.}{2021}]{lake2021}
{Lake} W.,  {Naoz} S.,  {Chiou} Y.~S.,  {Burkhart} B.,  {Marinacci} F.,
  {Vogelsberger} M.,   {Kremer} K.,  2021, \mn@doi [\apj]
  {10.3847/1538-4357/ac20d0}, \href
  {https://ui.adsabs.harvard.edu/abs/2021ApJ...922...86L} {922, 86}

\bibitem[\protect\citeauthoryear{{Lancaster}, {Ostriker}, {Kim}  \&
  {Kim}}{{Lancaster} et~al.}{2021a}]{lancaster2021a}
{Lancaster} L.,  {Ostriker} E.~C.,  {Kim} J.-G.,   {Kim} C.-G.,  2021a, \mn@doi
  [\apj] {10.3847/1538-4357/abf8ab}, \href
  {https://ui.adsabs.harvard.edu/abs/2021ApJ...914...89L} {914, 89}

\bibitem[\protect\citeauthoryear{{Lancaster}, {Ostriker}, {Kim}  \&
  {Kim}}{{Lancaster} et~al.}{2021b}]{lancaster2021b}
{Lancaster} L.,  {Ostriker} E.~C.,  {Kim} J.-G.,   {Kim} C.-G.,  2021b, \mn@doi
  [\apj] {10.3847/1538-4357/abf8ac}, \href
  {https://ui.adsabs.harvard.edu/abs/2021ApJ...914...90L} {914, 90}

\bibitem[\protect\citeauthoryear{{Larsen}}{{Larsen}}{2017}]{larsen2017}
{Larsen} S.~S.,  2017, in {Charbonnel} C.,  {Nota} A.,  eds,  IAU Symposium
  Vol. 316, Formation, Evolution, and Survival of Massive Star Clusters. pp
  91--98, \mn@doi{10.1017/S1743921315010509}

\bibitem[\protect\citeauthoryear{{Larsen}, {Brodie}, {Forbes}  \&
  {Strader}}{{Larsen} et~al.}{2014}]{larsen2014}
{Larsen} S.~S.,  {Brodie} J.~P.,  {Forbes} D.~A.,   {Strader} J.,  2014,
  \mn@doi [\aap] {10.1051/0004-6361/201322672}, \href
  {https://ui.adsabs.harvard.edu/abs/2014A&A...565A..98L} {565, A98}

\bibitem[\protect\citeauthoryear{{Lee}, {Shin}  \& {Kim}}{{Lee}
  et~al.}{2021}]{lee2020}
{Lee} J.,  {Shin} E.-j.,   {Kim} J.-h.,  2021, \mn@doi [\apjl]
  {10.3847/2041-8213/ac16e0}, \href
  {https://ui.adsabs.harvard.edu/abs/2021ApJ...917L..15L} {917, L15}

\bibitem[\protect\citeauthoryear{{Leitherer} et~al.,}{{Leitherer}
  et~al.}{1999}]{leitherer1999}
{Leitherer} C.,  et~al., 1999, \mn@doi [\apjs] {10.1086/313233}, \href
  {https://ui.adsabs.harvard.edu/abs/1999ApJS..123....3L} {123, 3}

\bibitem[\protect\citeauthoryear{{Li}, {Gnedin}, {Gnedin}, {Meng}, {Semenov}
  \& {Kravtsov}}{{Li} et~al.}{2017}]{li2017}
{Li} H.,  {Gnedin} O.~Y.,  {Gnedin} N.~Y.,  {Meng} X.,  {Semenov} V.~A.,
  {Kravtsov} A.~V.,  2017, \mn@doi [\apj] {10.3847/1538-4357/834/1/69}, \href
  {https://ui.adsabs.harvard.edu/abs/2017ApJ...834...69L} {834, 69}

\bibitem[\protect\citeauthoryear{{Li}, {Vogelsberger}, {Marinacci}  \&
  {Gnedin}}{{Li} et~al.}{2019}]{li2019a}
{Li} H.,  {Vogelsberger} M.,  {Marinacci} F.,   {Gnedin} O.~Y.,  2019, \mn@doi
  [\mnras] {10.1093/mnras/stz1271}, \href
  {https://ui.adsabs.harvard.edu/abs/2019MNRAS.487..364L} {487, 364}

\bibitem[\protect\citeauthoryear{{Li}, {Vogelsberger}, {Bryan}, {Marinacci},
  {Sales}  \& {Torrey}}{{Li} et~al.}{2021}]{li2021}
{Li} H.,  {Vogelsberger} M.,  {Bryan} G.~L.,  {Marinacci} F.,  {Sales} L.~V.,
  {Torrey} P.,  2021, arXiv e-prints, \href
  {https://ui.adsabs.harvard.edu/abs/2021arXiv210910356L} {p. arXiv:2109.10356}

\bibitem[\protect\citeauthoryear{{Ma} et~al.,}{{Ma} et~al.}{2020}]{Ma2020}
{Ma} X.,  et~al., 2020, \mn@doi [\mnras] {10.1093/mnras/staa527}, \href
  {https://ui.adsabs.harvard.edu/abs/2020MNRAS.493.4315M} {493, 4315}

\bibitem[\protect\citeauthoryear{{Madau}, {Lupi}, {Diemand}, {Burkert}  \&
  {Lin}}{{Madau} et~al.}{2020}]{madau2020}
{Madau} P.,  {Lupi} A.,  {Diemand} J.,  {Burkert} A.,   {Lin} D. N.~C.,  2020,
  \mn@doi [\apj] {10.3847/1538-4357/ab66c6}, \href
  {https://ui.adsabs.harvard.edu/abs/2020ApJ...890...18M} {890, 18}

\bibitem[\protect\citeauthoryear{{Mandelker}, {van Dokkum}, {Brodie}, {van den
  Bosch}  \& {Ceverino}}{{Mandelker} et~al.}{2018}]{mandelker2018}
{Mandelker} N.,  {van Dokkum} P.~G.,  {Brodie} J.~P.,  {van den Bosch} F.~C.,
  {Ceverino} D.,  2018, \mn@doi [\apj] {10.3847/1538-4357/aaca98}, \href
  {https://ui.adsabs.harvard.edu/abs/2018ApJ...861..148M} {861, 148}

\bibitem[\protect\citeauthoryear{{Martin} et~al.,}{{Martin}
  et~al.}{2022}]{martin2022}
{Martin} N.~F.,  et~al., 2022, arXiv e-prints, \href
  {https://ui.adsabs.harvard.edu/abs/2022arXiv220101310M} {p. arXiv:2201.01310}

\bibitem[\protect\citeauthoryear{{Martizzi}, {Faucher-Gigu{\`e}re}  \&
  {Quataert}}{{Martizzi} et~al.}{2015}]{martizzi2015}
{Martizzi} D.,  {Faucher-Gigu{\`e}re} C.-A.,   {Quataert} E.,  2015, \mn@doi
  [\mnras] {10.1093/mnras/stv562}, \href
  {https://ui.adsabs.harvard.edu/abs/2015MNRAS.450..504M} {450, 504}

\bibitem[\protect\citeauthoryear{{Mashchenko} \& {Sills}}{{Mashchenko} \&
  {Sills}}{2005}]{mashchenko2005}
{Mashchenko} S.,  {Sills} A.,  2005, \mn@doi [\apj] {10.1086/426133}, \href
  {https://ui.adsabs.harvard.edu/abs/2005ApJ...619..258M} {619, 258}

\bibitem[\protect\citeauthoryear{{Moore}, {Diemand}, {Madau}, {Zemp}  \&
  {Stadel}}{{Moore} et~al.}{2006}]{moore2006}
{Moore} B.,  {Diemand} J.,  {Madau} P.,  {Zemp} M.,   {Stadel} J.,  2006,
  \mn@doi [\mnras] {10.1111/j.1365-2966.2006.10116.x}, \href
  {https://ui.adsabs.harvard.edu/abs/2006MNRAS.368..563M} {368, 563}

\bibitem[\protect\citeauthoryear{{Murray}, {Quataert}  \& {Thompson}}{{Murray}
  et~al.}{2010}]{murray2010}
{Murray} N.,  {Quataert} E.,   {Thompson} T.~A.,  2010, \mn@doi [\apj]
  {10.1088/0004-637X/709/1/191}, \href
  {https://ui.adsabs.harvard.edu/abs/2010ApJ...709..191M} {709, 191}

\bibitem[\protect\citeauthoryear{{O{\~n}orbe}, {Garrison-Kimmel}, {Maller},
  {Bullock}, {Rocha}  \& {Hahn}}{{O{\~n}orbe} et~al.}{2014}]{onorbe2014}
{O{\~n}orbe} J.,  {Garrison-Kimmel} S.,  {Maller} A.~H.,  {Bullock} J.~S.,
  {Rocha} M.,   {Hahn} O.,  2014, \mn@doi [\mnras] {10.1093/mnras/stt2020},
  \href {https://ui.adsabs.harvard.edu/abs/2014MNRAS.437.1894O} {437, 1894}

\bibitem[\protect\citeauthoryear{{Pe{\~n}arrubia}, {Varri}, {Breen}, {Ferguson}
   \& {S{\'a}nchez-Janssen}}{{Pe{\~n}arrubia} et~al.}{2017}]{penarrubia2017}
{Pe{\~n}arrubia} J.,  {Varri} A.~L.,  {Breen} P.~G.,  {Ferguson} A. M.~N.,
  {S{\'a}nchez-Janssen} R.,  2017, \mn@doi [\mnras] {10.1093/mnrasl/slx094},
  \href {https://ui.adsabs.harvard.edu/abs/2017MNRAS.471L..31P} {471, L31}

\bibitem[\protect\citeauthoryear{{Peebles}}{{Peebles}}{1984}]{peebles1984}
{Peebles} P.~J.~E.,  1984, \mn@doi [\apj] {10.1086/161714}, \href
  {https://ui.adsabs.harvard.edu/abs/1984ApJ...277..470P} {277, 470}

\bibitem[\protect\citeauthoryear{{Peebles} \& {Dicke}}{{Peebles} \&
  {Dicke}}{1968}]{peebles1968}
{Peebles} P.~J.~E.,  {Dicke} R.~H.,  1968, \mn@doi [\apj] {10.1086/149811},
  \href {https://ui.adsabs.harvard.edu/abs/1968ApJ...154..891P} {154, 891}

\bibitem[\protect\citeauthoryear{{Pfeffer}, {Kruijssen}, {Crain}  \&
  {Bastian}}{{Pfeffer} et~al.}{2018}]{pfeffer2018}
{Pfeffer} J.,  {Kruijssen} J.~M.~D.,  {Crain} R.~A.,   {Bastian} N.,  2018,
  \mn@doi [\mnras] {10.1093/mnras/stx3124}, \href
  {https://ui.adsabs.harvard.edu/abs/2018MNRAS.475.4309P} {475, 4309}

\bibitem[\protect\citeauthoryear{{Phipps}, {Khochfar}, {Varri}  \& {Dalla
  Vecchia}}{{Phipps} et~al.}{2020}]{phipps2020}
{Phipps} F.,  {Khochfar} S.,  {Varri} A.~L.,   {Dalla Vecchia} C.,  2020,
  \mn@doi [\aap] {10.1051/0004-6361/202037884}, \href
  {https://ui.adsabs.harvard.edu/abs/2020A&A...641A.132P} {641, A132}

\bibitem[\protect\citeauthoryear{{Planck Collaboration} et~al.,}{{Planck
  Collaboration} et~al.}{2018}]{planck2018}
{Planck Collaboration} et~al., 2018, arXiv e-prints, \href
  {https://ui.adsabs.harvard.edu/abs/2018arXiv180706209P} {p. arXiv:1807.06209}

\bibitem[\protect\citeauthoryear{{Pozzetti}, {Maraston}  \&
  {Renzini}}{{Pozzetti} et~al.}{2019}]{pozzetti2019}
{Pozzetti} L.,  {Maraston} C.,   {Renzini} A.,  2019, \mn@doi [\mnras]
  {10.1093/mnras/stz785}, \href
  {https://ui.adsabs.harvard.edu/abs/2019MNRAS.485.5861P} {485, 5861}

\bibitem[\protect\citeauthoryear{{Reina-Campos}, {Kruijssen}, {Pfeffer},
  {Bastian}  \& {Crain}}{{Reina-Campos} et~al.}{2019}]{reina-campos2019}
{Reina-Campos} M.,  {Kruijssen} J.~M.~D.,  {Pfeffer} J.~L.,  {Bastian} N.,
  {Crain} R.~A.,  2019, \mn@doi [\mnras] {10.1093/mnras/stz1236}, \href
  {https://ui.adsabs.harvard.edu/abs/2019MNRAS.486.5838R} {486, 5838}

\bibitem[\protect\citeauthoryear{{Reina-Campos}, {Keller}, {Kruijssen},
  {Gensior}, {Trujillo-Gomez}, {Jeffreson}, {Pfeffer}  \&
  {Sills}}{{Reina-Campos} et~al.}{2022}]{reina-campos2022}
{Reina-Campos} M.,  {Keller} B.~W.,  {Kruijssen} J.~M.~D.,  {Gensior} J.,
  {Trujillo-Gomez} S.,  {Jeffreson} S. M.~R.,  {Pfeffer} J.~L.,   {Sills} A.,
  2022, arXiv e-prints, \href
  {https://ui.adsabs.harvard.edu/abs/2022arXiv220206961R} {p. arXiv:2202.06961}

\bibitem[\protect\citeauthoryear{{Renaud}}{{Renaud}}{2018}]{renaud2018}
{Renaud} F.,  2018, \mn@doi [\nar] {10.1016/j.newar.2018.03.001}, \href
  {https://ui.adsabs.harvard.edu/abs/2018NewAR..81....1R} {81, 1}

\bibitem[\protect\citeauthoryear{{Renaud}, {Agertz}  \& {Gieles}}{{Renaud}
  et~al.}{2017}]{renaud2017}
{Renaud} F.,  {Agertz} O.,   {Gieles} M.,  2017, \mn@doi [\mnras]
  {10.1093/mnras/stw2969}, \href
  {https://ui.adsabs.harvard.edu/abs/2017MNRAS.465.3622R} {465, 3622}

\bibitem[\protect\citeauthoryear{{Renzini}}{{Renzini}}{2017}]{renzini2017}
{Renzini} A.,  2017, \mn@doi [\mnras] {10.1093/mnrasl/slx057}, \href
  {https://ui.adsabs.harvard.edu/abs/2017MNRAS.469L..63R} {469, L63}

\bibitem[\protect\citeauthoryear{{Ricotti} \& {Gnedin}}{{Ricotti} \&
  {Gnedin}}{2005}]{ricotti2005}
{Ricotti} M.,  {Gnedin} N.~Y.,  2005, \mn@doi [\apj] {10.1086/431415}, \href
  {https://ui.adsabs.harvard.edu/abs/2005ApJ...629..259R} {629, 259}

\bibitem[\protect\citeauthoryear{{Ricotti}, {Parry}  \& {Gnedin}}{{Ricotti}
  et~al.}{2016}]{ricotti2016}
{Ricotti} M.,  {Parry} O.~H.,   {Gnedin} N.~Y.,  2016, \mn@doi [\apj]
  {10.3847/0004-637X/831/2/204}, \href
  {https://ui.adsabs.harvard.edu/abs/2016ApJ...831..204R} {831, 204}

\bibitem[\protect\citeauthoryear{{Schauer}, {Bromm}, {Boylan-Kolchin}, {Glover}
   \& {Klessen}}{{Schauer} et~al.}{2021}]{schauer2021}
{Schauer} A. T.~P.,  {Bromm} V.,  {Boylan-Kolchin} M.,  {Glover} S. C.~O.,
  {Klessen} R.~S.,  2021, \mn@doi [\apj] {10.3847/1538-4357/ac27aa}, \href
  {https://ui.adsabs.harvard.edu/abs/2021ApJ...922..193S} {922, 193}

\bibitem[\protect\citeauthoryear{{Simon} et~al.,}{{Simon}
  et~al.}{2021}]{simon2021}
{Simon} J.~D.,  et~al., 2021, \mn@doi [\apj] {10.3847/1538-4357/abd31b}, \href
  {https://ui.adsabs.harvard.edu/abs/2021ApJ...908...18S} {908, 18}

\bibitem[\protect\citeauthoryear{{Somerville}}{{Somerville}}{2002}]{somerville2002}
{Somerville} R.~S.,  2002, \mn@doi [\apjl] {10.1086/341444}, \href
  {https://ui.adsabs.harvard.edu/abs/2002ApJ...572L..23S} {572, L23}

\bibitem[\protect\citeauthoryear{{Sparre}, {Hayward}, {Feldmann},
  {Faucher-Gigu{\`e}re}, {Muratov}, {Kere{\v{s}}}  \& {Hopkins}}{{Sparre}
  et~al.}{2017}]{sparre2017}
{Sparre} M.,  {Hayward} C.~C.,  {Feldmann} R.,  {Faucher-Gigu{\`e}re} C.-A.,
  {Muratov} A.~L.,  {Kere{\v{s}}} D.,   {Hopkins} P.~F.,  2017, \mn@doi
  [\mnras] {10.1093/mnras/stw3011}, \href
  {https://ui.adsabs.harvard.edu/abs/2017MNRAS.466...88S} {466, 88}

\bibitem[\protect\citeauthoryear{{Spitler} \& {Forbes}}{{Spitler} \&
  {Forbes}}{2009}]{spitler2009}
{Spitler} L.~R.,  {Forbes} D.~A.,  2009, \mn@doi [\mnras]
  {10.1111/j.1745-3933.2008.00567.x}, \href
  {https://ui.adsabs.harvard.edu/abs/2009MNRAS.392L...1S} {392, L1}

\bibitem[\protect\citeauthoryear{{Springel} et~al.,}{{Springel}
  et~al.}{2008}]{springel2008}
{Springel} V.,  et~al., 2008, \mn@doi [\mnras]
  {10.1111/j.1365-2966.2008.14066.x}, \href
  {https://ui.adsabs.harvard.edu/abs/2008MNRAS.391.1685S} {391, 1685}

\bibitem[\protect\citeauthoryear{{Stinson}, {Dalcanton}, {Quinn}, {Kaufmann}
  \& {Wadsley}}{{Stinson} et~al.}{2007}]{stinson2007}
{Stinson} G.~S.,  {Dalcanton} J.~J.,  {Quinn} T.,  {Kaufmann} T.,   {Wadsley}
  J.,  2007, \mn@doi [\apj] {10.1086/520504}, \href
  {https://ui.adsabs.harvard.edu/abs/2007ApJ...667..170S} {667, 170}

\bibitem[\protect\citeauthoryear{Towns et~al.,}{Towns et~al.}{2014}]{xsede}
Towns J.,  et~al., 2014, \mn@doi [Computing in Science \& Engineering]
  {10.1109/MCSE.2014.80}, 16, 62

\bibitem[\protect\citeauthoryear{{Trenti}, {Padoan}  \& {Jimenez}}{{Trenti}
  et~al.}{2015}]{trenti2015}
{Trenti} M.,  {Padoan} P.,   {Jimenez} R.,  2015, \mn@doi [\apjl]
  {10.1088/2041-8205/808/2/L35}, \href
  {https://ui.adsabs.harvard.edu/abs/2015ApJ...808L..35T} {808, L35}

\bibitem[\protect\citeauthoryear{{Valenzuela}, {Moster}, {Remus}, {O'Leary}  \&
  {Burkert}}{{Valenzuela} et~al.}{2021}]{valenzuela2021}
{Valenzuela} L.~M.,  {Moster} B.~P.,  {Remus} R.-S.,  {O'Leary} J.~A.,
  {Burkert} A.,  2021, \mn@doi [\mnras] {10.1093/mnras/stab1701}, \href
  {https://ui.adsabs.harvard.edu/abs/2021MNRAS.505.5815V} {505, 5815}

\bibitem[\protect\citeauthoryear{{Vanzella} et~al.,}{{Vanzella}
  et~al.}{2017}]{vanzella2017}
{Vanzella} E.,  et~al., 2017, \mn@doi [\mnras] {10.1093/mnras/stx351}, \href
  {https://ui.adsabs.harvard.edu/abs/2017MNRAS.467.4304V} {467, 4304}

\bibitem[\protect\citeauthoryear{{Vanzella} et~al.,}{{Vanzella}
  et~al.}{2019}]{vanzella2019}
{Vanzella} E.,  et~al., 2019, \mn@doi [\mnras] {10.1093/mnras/sty3311}, \href
  {https://ui.adsabs.harvard.edu/abs/2019MNRAS.483.3618V} {483, 3618}

\bibitem[\protect\citeauthoryear{{Vanzella} et~al.,}{{Vanzella}
  et~al.}{2021a}]{vanzella2021}
{Vanzella} E.,  et~al., 2021a, arXiv e-prints, \href
  {https://ui.adsabs.harvard.edu/abs/2021arXiv210610280V} {p. arXiv:2106.10280}

\bibitem[\protect\citeauthoryear{{Vanzella} et~al.,}{{Vanzella}
  et~al.}{2021b}]{vanzella2021b}
{Vanzella} E.,  et~al., 2021b, \mn@doi [\aap] {10.1051/0004-6361/202039466},
  \href {https://ui.adsabs.harvard.edu/abs/2021A&A...646A..57V} {646, A57}

\bibitem[\protect\citeauthoryear{{Virtanen} et~al.,}{{Virtanen}
  et~al.}{2020}]{scipy2020}
{Virtanen} P.,  et~al., 2020, \mn@doi [Nature Methods]
  {10.1038/s41592-019-0686-2}, \href
  {https://ui.adsabs.harvard.edu/abs/2020NatMe..17..261V} {17, 261}

\bibitem[\protect\citeauthoryear{{Walch} \& {Naab}}{{Walch} \&
  {Naab}}{2015}]{walch2015}
{Walch} S.,  {Naab} T.,  2015, \mn@doi [\mnras] {10.1093/mnras/stv1155}, \href
  {https://ui.adsabs.harvard.edu/abs/2015MNRAS.451.2757W} {451, 2757}

\bibitem[\protect\citeauthoryear{{Wetzel} \& {Garrison-Kimmel}}{{Wetzel} \&
  {Garrison-Kimmel}}{2020a}]{wetzel2020b}
{Wetzel} A.,  {Garrison-Kimmel} S.,  2020a, {HaloAnalysis: Read and analyze
  halo catalogs and merger trees} (\mn@eprint {ascl} {2002.014})

\bibitem[\protect\citeauthoryear{{Wetzel} \& {Garrison-Kimmel}}{{Wetzel} \&
  {Garrison-Kimmel}}{2020b}]{wetzel2020}
{Wetzel} A.,  {Garrison-Kimmel} S.,  2020b, {GizmoAnalysis: Read and analyze
  Gizmo simulations} (\mn@eprint {ascl} {2002.015})

\bibitem[\protect\citeauthoryear{{Wheeler} et~al.,}{{Wheeler}
  et~al.}{2019}]{wheeler2019}
{Wheeler} C.,  et~al., 2019, \mn@doi [\mnras] {10.1093/mnras/stz2887}, \href
  {https://ui.adsabs.harvard.edu/abs/2019MNRAS.490.4447W} {490, 4447}

\bibitem[\protect\citeauthoryear{{Zick}, {Weisz}  \& {Boylan-Kolchin}}{{Zick}
  et~al.}{2018}]{zick2018}
{Zick} T.~O.,  {Weisz} D.~R.,   {Boylan-Kolchin} M.,  2018, \mn@doi [\mnras]
  {10.1093/mnras/sty662}, \href
  {https://ui.adsabs.harvard.edu/abs/2018MNRAS.477..480Z} {477, 480}

\makeatother
\end{thebibliography}




\bsp	
\label{lastpage}
\end{document}